\documentclass{aastex631}

\usepackage{bm}
\usepackage{subcaption}
\usepackage[T1]{fontenc}
\usepackage{booktabs}
\usepackage{threeparttable}
\usepackage{multirow}
\usepackage{amsmath}
\usepackage[utf8]{inputenc}
\usepackage{makecell}
\usepackage{hyperref}
\usepackage{numprint}
\usepackage[ruled,vlined]{algorithm2e}
\usepackage{tabularx}

\begin{document}

 \title{CosForce: A Force-Based General Pedestrian Model with Anticipation and Reaction Mechanisms}
 
\author{Jinghui Wang}

\affiliation{School of Safety Science and Emergency Management\\
Wuhan University of Technology\\
Wuhan, China}

\author{Wei Lv}
\altaffiliation{{\url{weil@whut.edu.cn} (W. Lv)}}
\affiliation{School of Safety Science and Emergency Management\\
Wuhan University of Technology\\
Wuhan, China}
\affiliation{China Research Center for Emergency Management\\
Wuhan University of Technology\\
Wuhan, China}

\author{Shuchao Cao}

\affiliation{School of Automotive and Traffic Engineering\\
Jiangsu University\\
Zhenjiang, China}

\author{Chenglin Guo}

\affiliation{State Key Laboratory of Fire Science\\
University of Science and Technology of China\\
Hefei, China}

\begin{abstract}
This paper develops a general force-based pedestrian model named CosForce, in which cosine functions are employed to describe asymmetric interactions. These functions implicitly capture the mechanisms of anticipation and reaction. By focusing on binary interactions, the CosForce model provides new insights into pedestrian modeling while achieving property of linear time complexity. Two specific scenarios in crowd dynamics were analyzed: self-organization (entropy decrease) and crowd collapse (entropy increase). The average normalized speed and order parameter were introduced to quantitatively describe the processes of crowd dynamics. Quantitative evaluations demonstrate that phase separation in crowds is effectively reproduced by the model, including lane formation, stripe formation, and cross-channel formation. Next, in the simulation of mass gathering, within a density-accumulating scenario, processes of critical transition in high-density crowds are clearly revealed through time series observations of the order parameter. These findings provide valuable insights into crowd dynamics.
\end{abstract}






\section{Introduction} 
\label{section1}

The study of crowd dynamics has evolved over several decades. Although pedestrian dynamics are closely related to daily lives, our understanding of crowd behavior remains incomplete. As a field at the intersection of traffic dynamics and collective dynamics, it has naturally integrated perspectives and theories from various disciplines, such as theory of  pedestrian fundamental diagram \citep{seyfried2005fundamental,zhang2012ordering} and critical transition mechanisms \citep{scheffer2012anticipating, vicsek1995novel, szabo2006phase}, among others. Field studies of Hajj \citep{helbing2007dynamics}, the Love Parade disaster \citep{ma2013new}, and the San Fermín festival \citep{gu2024emergence} have demonstrated the fluid-like properties of dense crowds. These findings indicate that zero-flow density does not strictly exist in crowd systems, establishing a new paradigm for crowd analysis. However, a comprehensive theoretical framework that fully captures crowd dynamics is still lacking. Empirical studies provide extensive data on pedestrian behavior under normal conditions, yet precise data for extreme scenarios remain scarce \citep{haghani2018crowd}. Moreover, due to the property of self-propelled and social nature of crowds, behavioral diversity makes it challenging to establish universally applicable observations. These difficulties complicate the simplification of crowd movement into a straightforward particle-interaction process for analysis, yet such an approach remains essential.

Microscopic pedestrian dynamics models mainly focus on real-time interactions based on operational-level simulation, such as the cellular automaton model (CA) \citep{burstedde2001simulation}, the force-based model \citep{helbing1995social, chraibi2010generalized}, the collision avoidance model \citep{van2008reciprocal, tordeux2016collision}, and the heuristic model \citep{moussaid2011simple}, have been widely applied. Researchers have also sought inspiration from interdisciplinary knowledge, such as the concept of order parameters derived from Landau theory, the mean-field game theory \citep{bonnemain2023pedestrians}, and the PLE model \citep{guy2010pledestrians} based on Principle of Least Action. Based on the memoryless property of Markov processes, physics-based approaches incorporate stochastic noise to avoid deterministic simulation. Another category, trajectory prediction methods, from a statistical perspective, where time-series inference is elevated to a paramount position \citep{alahi2016social}. Numerical validation results suggest that data-driven models may surpass physics-based approaches, at least in the tasks of short-term pedestrian trajectory prediction (a few seconds, low-density). Nevertheless, in analyzing crowd dynamics, physics-based modeling remains the most reliable approach as it operationalizes the core principles of human reasoning.

Under normal conditions, pedestrians exhibit noncontact interactions, where collision avoidance behavior is constrained by empirical relationships. The most commonly used principle is the linear relationship between headway (which can be approximated in a 2D space by the distance with nearest neighbor in the direction of motion) and speed. Researchers can add new rules to investigate specific mechanisms \citep{liang2021continuum,shang2024development}. Among these, anticipation and reaction have been extensively analyzed. Collision avoidance rules can be introduced as modeling based on anticipation behavior \citep{zanlungo2011social, gerlee2017impact, lu2020pedestrian, hu2024crowd, xu2021anticipation}. In extreme density, the collision avoidance mechanisms of pedestrians will completely fail, and Newton’s third law becomes the fundamental principle governing crowd dynamics. The motion of the crowd therefore displays properties similar to particle hydrodynamics \citep{van2020extreme}. Force-based models provide a comprehensive physical framework for particle simulation, with the capability to accurately model interactions within dense crowds. Among these, a potential issue is that the complex models often perform poorly in cross-scenario simulations. Such models are frequently effective in specific tasks; however, the optimization for particular objectives may, in fact, undermine their generalization capability. More importantly, the cumbersome rules and formulas limit the ability of model for provide straightforward insight in pedestrian dynamics. Furthermore, Simulation results are challenging to validate objectively. In this regard, it may be more effective to develop a general model by simplifying models rather than by adding complexity. Everyone naturally engages with motion in their daily lives. Therefore, preserving empirical insights and intuitive understanding is essential. This principle, as discussed below, forms the cornerstone of the model proposed in this paper.

\textbullet\ \textbf{Anisotropy of pedestrian motion space}

Traditionally, the space surrounding pedestrians has been assumed to be isotropic. This perspective leads to a potential modeling issue: forces from different directions may produce opposing effects, causing pedestrians to maintain considerable speeds in crowded situations. Such results contradict empirical observations of fundamental diagrams in pedestrian dynamics \citep{parisi2009modification}. Furthermore, similar models have yet to attain optimal accuracy in crowd modeling, as they demonstrate effectiveness primarily in specific contexts, such as densely packed crowds. In such cases, vision-based or attention-driven motion patterns are ineffective, and direct physical contact leads the dense crowd to approximate the behavior of a granular system.

\begin{figure}[ht!]
\centering
\includegraphics[scale=1.1]{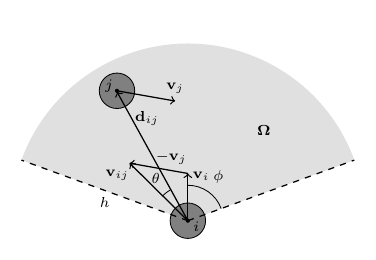}
\caption{Diagram of the process of pedestrian interaction.}
\label{fig1}
\end{figure}

Empirical observations of pedestrian behavior reveal that individuals primarily focus on the dynamics in front of them, largely disregarding those behind. Prior to introducing the model, the Horizontal Field of Attention (HFA, \(\mathbf{\Omega} \in {\mathbb{R}^2}\)) in our model is defined. The HFA is constrained by the attentional eccentricity angle \(\phi\) and the attentional depth \(h\), as depicted in Fig.\ref{fig1}. The angle \(\phi\) represents the angle between the boundary of the sector-shaped HFA and the pedestrian's direction. The attentional depth \(h\) can theoretically extend to infinity. To minimize computation, \(h\) can be set to the minimum headway allowing pedestrians to maintain free motion.

\textbullet\ \textbf{Pedestrians resemble compressible particles rather than rigid particles}

In the classical social force model, pedestrians are represented as rigid particles. In narrow passages, pedestrians frequently turn sideways, and mutual compression reduces the space occupied. In order to capture more detailed motion dynamics, some studies have proposed the use of ellipsoidal shapes instead of circular ones to simulate pedestrian movement \citep{chraibi2010generalized}. On the other hand, this modification will leads to an increase in model complexity. An effective compromise can be achieved by modeling pedestrians as compressible bodies allows for a more concise simulation of these behaviors \citep{narain2009aggregate}. The fluid-like properties observed in crowd dynamics have led researchers to adopt hydrodynamic perspective for crowd analysis. Empirical studies already demonstrated the presence of non-zero divergence in crowds \citep{johansson2008crowd}, providing substantial evidence that supports the conceptualization of pedestrian crowds as compressible media.

\begin{figure}[ht!]
\centering
\includegraphics[scale=0.58]{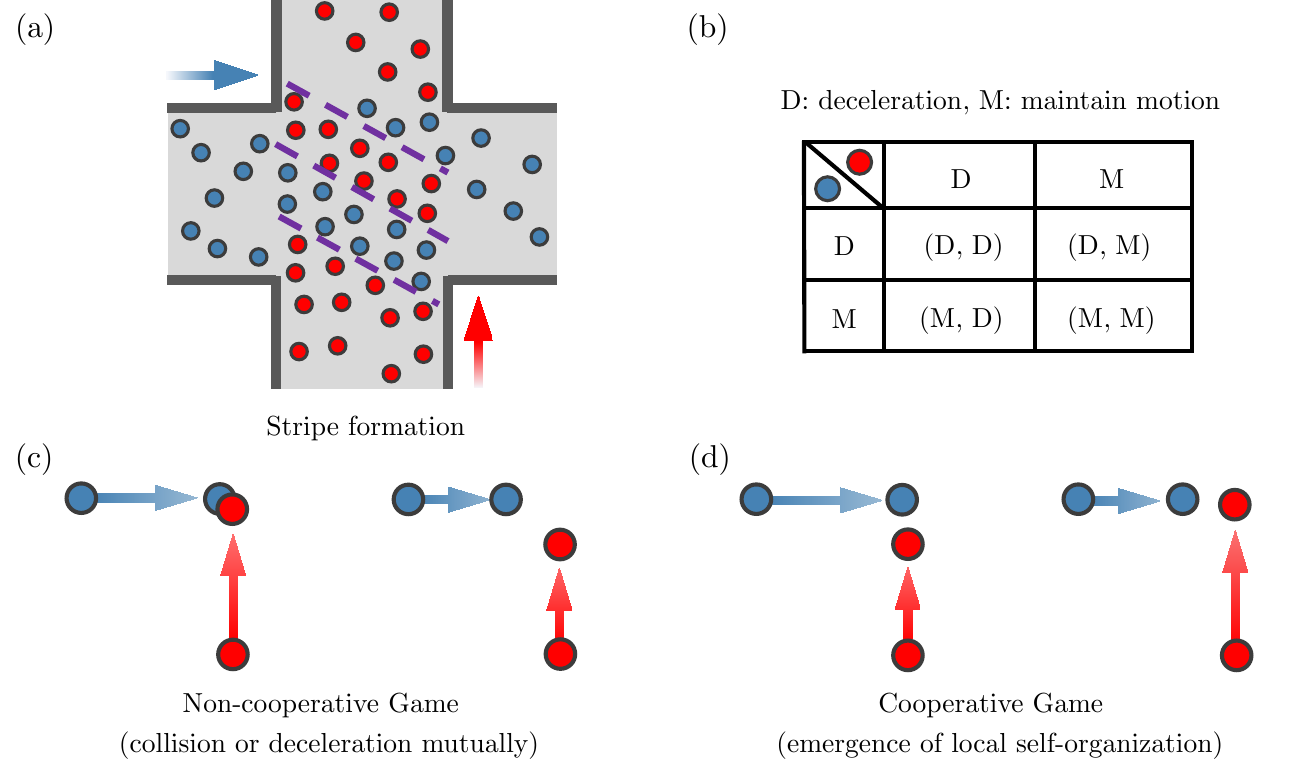}
\caption{Scaling down the laws of 
 stripe formation. (a) Schematic diagram of stripe formation. (b) Decision payoff matrix for pedestrians in binary conflicts. (c) Cooperative and non-cooperative games between conflicting pedestrians.}
\label{fig2}
\end{figure}

\textbullet\ \textbf{Collisions are necessary}

The models based on collision avoidance algorithms has proven to be highly effective in studying interactions within crowds and is currently widely used in modeling 2-D multi-agent systems for obstacle avoidance \citep{van2008reciprocal,tordeux2016collision}. A potential issue is that the precise collision avoidance algorithm actually limits its capability for high-accuracy crowd simulation: in crowds, collisions are common and, in some situations, even inevitable. This discrepancy arises mainly because pedestrian speed and orientation are updated based on empirical estimates rather than precise numerical calculations. In reality, pedestrians can not perform such calculations, their motion mainly rely on approximations. In this consideration, the collision process is crucial for pedestrian simulation, especially in dense crowds.

\textbullet\ \textbf{How many pedestrians can we tracked simultaneously?}

Research based on collective dynamics typically establishes models from a bottom-up perspective, wherein micro-level rules govern the evolution of groups, such as the BOID model \citep{reynolds1987flocks} and network models \citep{bode2011social,allen2017evolutionary}. Based on metrics \citep{vicsek1995novel}, topological \citep{ballerini2008interaction}, or visual connection \citep{rosenthal2015revealing, wirth2023neighborhood}, the state of individuals exhibits a strong correlation with the characteristics of their surrounding neighborhoods (coherence of motion). The complexity and variability of pedestrian dynamics hinder the approximation of crowds to fish schools or flocks of birds. However, the paradigms of interactions observed can indeed be transferred across species.

When pedestrians move within a crowd, they primarily acquire information based on vision. The visual field of pedestrians encompasses a considerable range, potentially including dozens or even hundreds of pedestrians and possible obstacles. However, the range within which pedestrians actually focus their attention is quite limited. Similar mechanism has been observed in experimental research within the field of cognitive science. Interactive experiments has revealed that visual attention is deployed differentially, depending on the nature of the behavioral goal that designates the task relevance of visual input \citep{renton2019differential,frielink2017distinguishing}. In addition to the "magical number 4" theory \citep{cowan2001magical} concerning visual attention, mainstream perspectives suggest that the complexity and dynamism of the environment affect the number of observable entities, with higher levels of dynamism and complexity leading to a reduction in the number of objects that pedestrians can track \citep{alvarez2007many}.

In highly stochastic and dynamic crowds, pedestrians can realistically focus on only a limited number of targets, typically their nearest neighbors within their attention field. From this perspective, models that employ binary interactions rather than metric-based or pairwise interactions may be more concise and efficient.

\textbullet\ \textbf{Cooperative game induces local self-organization}

The phenomenon of self-organization among pedestrians is characterized by entropy decrease and the emergence of steady states within crowd systems. From a reductionist perspective, we conjecture that such entropy decrease is presumably induced by local cooperative interactions among conflicting pedestrians \citep{rand2014static,su2022evolution,bonnemain2023pedestrians, zablotsky2024pedestrian}. In a typical crossing flow scenario, as depicted in Fig.\ref{fig2}(a), there is a potential risk of collision between crossing pedestrians. Anticipation mechanism, involving a negotiation or game process, can facilitate "phase separation" between pedestrians moving in different directions. The interaction results between pedestrians from distinct directions can be represented by a simple payoff matrix, as shown in Fig.\ref{fig2}(b). Cooperative games contribute to entropy decrease within the system, leading to the emergence of spontaneous order, as illustrated in Fig.\ref{fig2}(d). In this consideration, the local cooperation rule is implicitly embedded in our model through asymmetric force interactions governed by the cosine function.

From the perspective of applicability and simplicity, the general pedestrian model of the operational layer should incorporate the following characteristics: \textbf{Empirical foundations}: grounded in empirical data and aligned with our experience; \textbf{Minimalist principles}: employing the simplest rules to reveal core mechanisms while allowing straightforward analysis; \textbf{Diversity in model performance}: parameters can transfer across scenarios, consistent with fundamental diagrams, diversity in simulation (e.g., phenomena of pedestrian self-organization); \textbf{Computational efficiency}: lower computational cost compared to other models in the same category. To this end, based on the above requirements, a general model of pedestrian dynamics is established in this paper. The rules of the model, developed from minimalism principles, were further refined in accordance with the classic social force model to enhance simulation accuracy. Additionally, this minimalist approach facilitates the analytical examination of pedestrian interactions. The subsequent sections of the paper unfolded as follows: In section \ref{section2}, considering the anticipation and reaction behavior of pedestrians, we established the CosForce model. In section \ref{section3}, empirical validations were conducted for single-file motion and unidirectional flow. Subsequently, to investigate two specific mechanisms in crowd dynamics, namely the spontaneous processes of entropy decrease and entropy increase, we introduced fundamental parameters for quantitative evaluation the self-organization processes in section \ref{section4} and the phenomenon of crowd collapse in section \ref{section5}. Based on the modeling assumption of binary interactions, we validated the advantage of linear time complexity in computational efficiency in section \ref{section6} and discussed the limitations. Finally, section \ref{section7} summarized the conclusions drawn from the study.

\section{Methodology} \label{section2}

\subsection{Constraint of Space-speed Relationship} \label{subsection2.1}

\begin{figure}[ht!]
\centering
\includegraphics[scale=0.75]{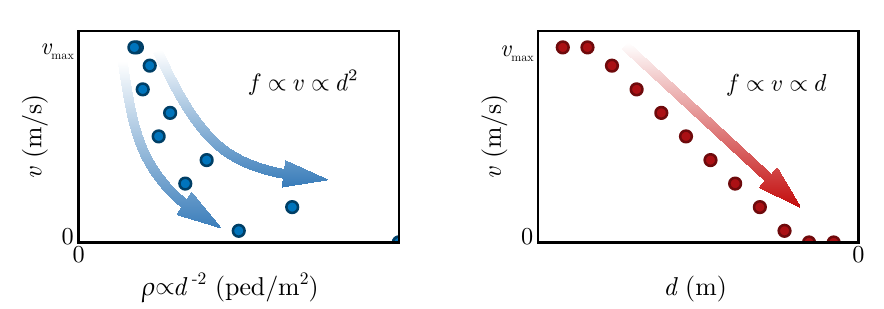}
\caption{Space-speed relationship. (a) Speed-local density constraint. (b) Speed-headway constraint.}
\label{fig3}
\end{figure}

The correlation between speed and space has been thoroughly investigated in experimental studies, as depicted in Fig.\ref{fig3}. Under the assumption of isotropic pedestrian interactions, it has been demonstrated that, in constrained scenarios, velocity is inversely correlated with local density (i.e., the speed–density relationship), as articulated in Eq.\ref{1}, with a constant parameter denoted as \(k\) (dimension specified as \( \mathrm{m}^{-1}\,\mathrm{s}^{-1} \)). This empirical relationship forms the foundation of force-based methods, treating the relaxation time as a constant, wherein the variation of pedestrian force with respect to spatial metrics is characterized by a power-law function, as detailed in Eq.\ref{2}.

\begin{equation}\label{1}
v \propto \rho^{-1} \propto d^2
\end{equation}

\begin{equation}\label{2}
\left\{
\begin{aligned}
\displaystyle \Delta v&= k\Delta d^2, \\
\displaystyle \Delta f&= m \frac{\Delta v}{\Delta t}= mk \frac{\Delta d^2}{\Delta t}.
\end{aligned}
\right.
\end{equation}

Furthermore, in one-dimensional motion, a linear empirical relationship between the headway and speed is typically employed, often associated with the principle of constant time headway. Considering the anisotropy of the pedestrian movement space, this linear relationship, based on the nearest neighbor distance, can similarly be extended to two-dimensional motion, as illustrated in Eq.\ref{3}. In this context, considering a constant relaxation time, a linear function can be employed to describe the variation of pedestrian force with respect to the relative distance to the nearest neighbor, as shown in Eq.\ref{4}.

\begin{equation}\label{3}
d = v t_h + d_0 \quad \Rightarrow  \quad v = \frac{d - d_0}{t_h}
\end{equation}

\begin{equation}\label{4}
\left\{
\begin{aligned}
\frac{\Delta d}{\Delta v} &= t_h, \\
\Delta f &=\frac{m \Delta v}{\Delta t} = \frac{m \Delta d}{t_h \, \Delta t}.
\end{aligned}
\right.
\end{equation}

Based on these property, it is natural to employ power functions or linear functions of distance to describe such mechanisms. The empirical formulas underlying linear functions demonstrates scale invariance, contributes to its stability.
Based on the hypothesis of binary interactions, each pedestrian is only subject to two forces under non-collision conditions: the self-driven force \(f_i\) and the repulsive force \(f_{ij}\) from nearest-neighbor. Consequently, the 1-D equilibrium properties of pedestrians can be derived as follows: 

\begin{equation}\label{5}
v_i \to \max \left( \min \left( \frac{{d_{ij} - r_{ij}}}{{t_h}}, v_{\max} \right), 0 \right) \Rightarrow f_{ij} \to f_i.
\end{equation}  

The scalar form of the self-driven force \(\mathbf{f}_i\) is defined by Eq.\ref{6}. Therefore, the equation for the nearest-neighbor repulsive force at equilibrium position can be  derived as shown in Eq.\ref{7}.

\begin{equation}\label{6}
{f}_{i} = \frac{{{m_i}}}{\tau }\left( {{v}_{\max} - {v}_{i}} \right)
\end{equation}

\begin{equation}\label{7}
{f}_{ij} = \frac{{{m_i}}}{\tau }\left( {{v}_{\max }} - \max \left( {\min \left( {\frac{{ {{d}_{ij}} - {r_{ij}}}}{{{t_h}}}, {{v}_{\max }}}\right),0} \right)\right)
\end{equation}

\subsection{Implicit Modeling of Anticipation and Reaction Mechanisms} \label{subsection2.2}

In force-based methods, explicitly modeling the mechanisms of anticipation and reaction is often cumbersome. First, the underlying dynamics of the force-based model already represent the reaction mechanism through velocity adaptation constrained by a constant relaxation time. On this basis, the anticipation mechanism related to collision avoidance requires taking the future state of neighboring pedestrians (at a anticipation time of \(\delta\)) as input. This process potentially involves strategic interactions that require specific treatments and are therefore difficult to generalize into uniform rules. Therefore, we adopt an implicit modeling approach for anticipation mechanisms, aiming for simplicity.

\textbullet\ \textbf{Collision avoidance based on cos\(\theta\)}

In the computation based on 2-D TTC, the angle \(\theta\) between the relative velocity and relative distance plays a critical role in collision avoidance processes (\(\cos \theta\) represents the projection between the relative velocity and relative distance). In our model, a scaling factor (\(1 + \alpha \cos \theta\)) is introduced into the nearest-neighbor repulsive force to describe the effect of collision avoidance. Here, \(\alpha \in [0,1]\) is a dimensionless coefficient that regulates the collision avoidance scale of pedestrians, satisfying \((1 + \alpha \cos \theta) \in [1 - \alpha, 1 + \alpha]\). \(\alpha \to 0\) indicates that pedestrians are insensitive to collisions, which typically corresponds to unidirectional flow. \(\alpha \to 1\) signifies that pedestrians are highly sensitive to collisions, corresponding to crowd or multi-directional flow. Based on the deduce, we derived the equation of the nearest-neighbor repulsive force as a function of distance \(d\) and angle \(\theta\), as shown in Eq.\ref{8}.

\begin{equation}\label{8}
{f}_{ij} = \frac{{{m_i}}}{\tau }\left( {{v}_{\max }}  - \max \left( {\min \left( {\frac{{ {{d}_{ij}} - {r_{ij}}}}{{{t_h}}}, {{v}_{\max }} } \right),0} \right)\right){\left( {1 + \alpha \cos \theta } \right)}
\end{equation}

\begin{figure}[ht!]
\centering
\includegraphics[scale=0.58]{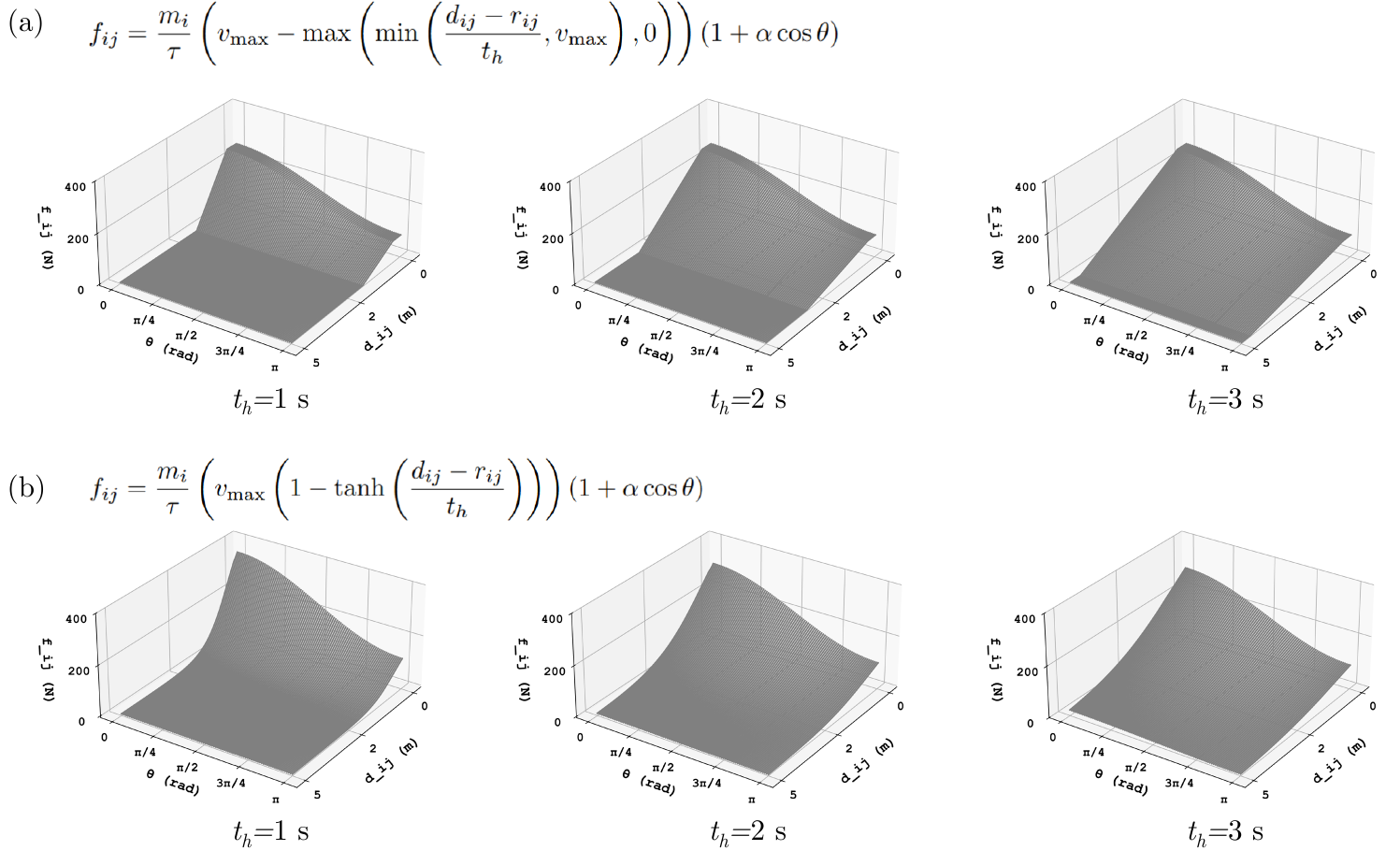}
\caption{Relationship between the repulsive force and $d$ and $\theta$, ($\alpha = 0.5$), (a) piecewise function, (b) continuous function.}
\label{fig4}
\end{figure}

For the simplicity, the tanh function can be employed to describe its continuous form, which aligns with the formulation used to represent the desired velocity in traffic flow. The continuous form is provided in Eq.\ref{9}. Fig.\ref{fig4} illustrated the influence of the nearest-neighbor relative distance \(d\) and angle \(\theta\) on the force term.

\begin{equation}\label{9}
{f}_{ij} = \frac{{m_i}}{\tau} \left( {v}_{\max}  \left( 1 - \tanh \left( \frac{ {d}_{ij}  - r_{ij}}{t_h} \right) \right) \right) \left( 1 + \alpha \cos \theta \right)
\end{equation}

\textbullet\ \textbf{Symmetric and asymmetric interactions}

\begin{figure}[ht!]
\centering
\includegraphics[scale=0.7]{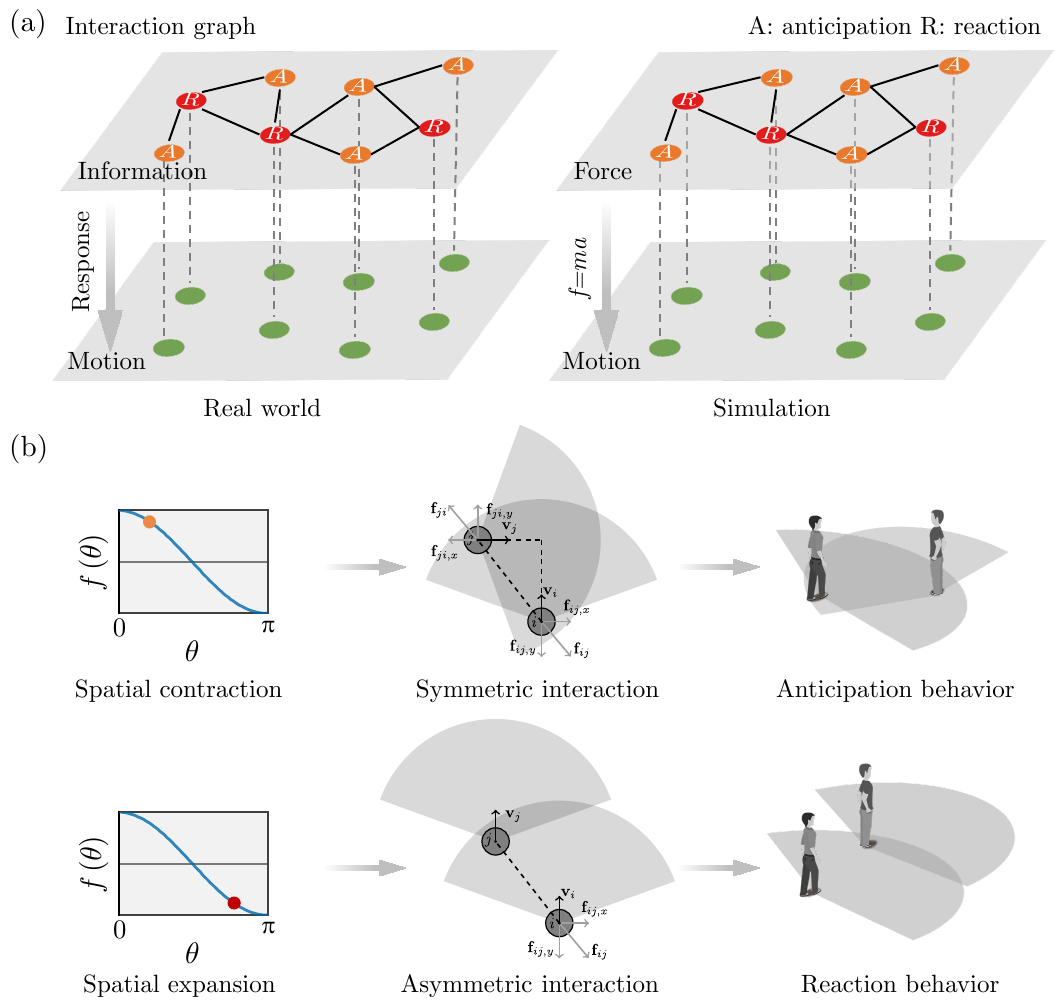}
\caption{Schematic of nearest neighbor interaction: (a) illustration of force-based pedestrian simulation, (b) implicit modeling of anticipation and reaction mechanisms through fine-tuning in force-based models.}
\label{fig5}
\end{figure}

Based on the HFA described above, pedestrian interactions involve both reciprocal and non-reciprocal interactions, as illustrated in Fig.\ref{fig5}. Symmetric and asymmetric forces naturally be employed to represent these mechanisms. Overall, these mechanisms are associated with the anticipation and reaction behaviors observed in pedestrian motion. In this context, the anticipation mechanism is referred to as a implicit cooperative game process among conflicting pedestrians (inducing separation). 
Conversely, the reaction mechanism is described as the convergence process among co-directional pedestrians (inducing aggregation).

\subsection{Model Rules} \label{subsection2.3}

Based on the framework of a force-based method, the model is characterized by a self-driven force \(\mathbf{f}_i\), a repulsive force \(\mathbf{f}_{ij}\) exerted by the nearest entity (pedestrian or wall \(j\), \(j \in \mathbf{\Omega} : = \lVert {{\mathbf{d}_{ij}}} \rVert < h \wedge \angle \left( {{\mathbf{v}_i},{\mathbf{d}_{ij}}} \right) < \phi \)), and the contact forces \(\mathbf{f}_c\) that arise during collisions. The global constraints can be expressed as follows:

\begin{equation}\label{10}
\mathbf{f} = \mathbf{f}_i + \mathbf{f}_{ij} 
+ \sum_{\lVert \mathbf{d}_{ij}\rVert < r_{ij}} \mathbf{f}_c = m\mathbf{a}
\end{equation}

Here, \(\mathbf{f}\) represents the net force acting on a target pedestrian, \(\mathbf{f}_{i}\) denotes the self-driven force of the pedestrian, \(\mathbf{f}_{ij}\) represents the repulsive force from the nearest entity \(j\).

The self-driven force is driven by the maximum velocity \(\mathbf{v}_{\text{max}}\), expressed as:

\begin{equation}\label{11}
\mathbf{f}_{i} = \frac{{{m_i}}}{\tau }\left( {\mathbf{v}_{\max} - \mathbf{v}_{i}} \right)
\end{equation}

Here, \(\tau\) represents the relaxation time, and \(m_i\) denotes the mass of pedestrian \(i\).

The interaction force between a pedestrian and the nearest entity \(j\) is represented by a repulsive force, which can be expressed in either piecewise or continuous forms as shown in Eq.\ref{12} and Eq.\ref{13}. In our simulation, the piecewise function was utilized to construct the repulsive forces.

Piecewise form:
\begin{equation}\label{12}
\mathbf{f}_{ij} = \frac{{{m_i}}}{\tau }\left(\lVert {\mathbf{v}_{\max }} \rVert - \max \left( {\min \left( {\frac{{\lVert \mathbf{d}_{ij} \rVert- {r_{ij}}}}{{{t_h}}},\lVert {\mathbf{v}_{\max }} \rVert} \right),0} \right)\right){\left( {1 + \alpha \cos \theta } \right)}\mathbf{n}_{ij}
\end{equation}

Continuous form:
\begin{equation}\label{13}
\mathbf{f}_{ij} = \frac{{m_i}}{\tau} \left( \lVert \mathbf{v}_{\max} \rVert \left( 1 - \tanh \left( \frac{\lVert \mathbf{d}_{ij} \rVert - r_{ij}}{t_h} \right) \right) \right) \left( 1 + \alpha \cos \theta \right) \mathbf{n}_{ij}
\end{equation}

Here, \( \theta \) is the angle between the relative distance and the relative velocity, \(\theta = \angle \left( \mathbf{v}_{ij}, \mathbf{d}_{ij} \right) \in \left[ 0, \pi \right]\). Other symbols are defined as follows: \(r_{ij} = r_i + r_j\), \(\mathbf{v}_{ij} = \mathbf{v}_i - \mathbf{v}_j\), and \(\mathbf{d}_{ij} = \mathbf{x}_j - \mathbf{x}_i\), \({\mathbf{n}_{ij}} =  - {\mathbf{d}_{ij}}/ \lVert {{\mathbf{d}_{ij}}} \rVert\). In this process, obstacles are modeled as stationary pedestrians and are therefore included in the calculation of the repulsive force without requiring any additions. Different from the point-to-point interaction between pedestrians, the interaction between pedestrians and obstacles follows a point-to-line relationship. Considering all potential collision scenarios, the attentional eccentricity angle $\phi$ for pedestrian-wall interaction is set to a constant value of $\pi/2$.

The collision force is described by an exponential decay term, which is activated when the condition \(\|\mathbf{d}_{ij}\| < r_{ij}\) is satisfied, as follows:

\begin{equation}\label{14}
\mathbf{f}_c = e^{\frac{r_{ij} - \lVert \mathbf{d}_{ij} \rVert}{\lambda}} \mathbf{n}_{ij}
\end{equation}

As mentioned earlier, the pedestrian is represented as a compressible particle, described by an exponential function. When the pedestrian's body undergoes compression, the resulting repulsive force increases exponentially with the compression magnitude. A fully repulsive core exists as the compression limit for the pedestrian particle. The parameter settings are presented in Tab.\ref{table1}, where \(m = 60\) kg, \(r = 0.2\) m, \(\tau = 0.5\) s, \(t_h = 1.3\) s, \(\lambda = 0.8\) and \(v_{\rm{max}}\)= 1.4 m/s are global parameters that remain constant across all simulations. The variable parameters \(\phi\), and \(\alpha\) will be adjusted as required. The frame rate for all simulations is set as 30, corresponding to a difference time of \(1/30\) s.

\begin{table}[htbp]
\centering
\caption{Parameter descriptions and reference values}
\label{table1} 
\begin{tabular}{cccc}
\toprule
\textbf{Parameter} & \textbf{Description} & \textbf{Type} & \textbf{Reference value} \\
\midrule
$m$ & Mass & Constant & 60 kg \\
$v_{\rm{max}}$ & Maximum speed & Constant & 1.4 m/s \\
$r$ & Radius & Constant & 0.2 m \\
$\tau$ & Relaxation time & Constant & 0.5 s \\
$\lambda$ & Dimensionless coefficient & Constant & 0.8 (scale in centimeters) \\
$t_h$ & Approximate time headway & Variable & 1.3 s \\
$\phi$ & Attentional eccentricity angle & Variable &  [0 - $\pi$] rad \\
$\alpha$ & Dimensionless coefficient & Variable & [0-1] \\
\bottomrule
\end{tabular}
\end{table}

\subsection{1-D dynamical functions of CosForce model} \label{subsection2.4}

\textbullet\ \textbf{Approximated 2-D FVD Model under Force-based Framework}

We have presented the 1-D form of the CosForce model to facilitate the analysis of the model's properties, the dynamical equation is expressed as:

\begin{equation}\label{15}
f = m_i a = \frac{m_i}{\tau} \left( \max \left( \min \left( \frac{ {d}_{ij}  - r_{ij}}{t_h},  {v}_{\max} \right), 0 \right) - {v_i} \right) \left( 1 + \alpha \, \operatorname{sgn} \left( {v}_{ij}  \right) \right)
\end{equation}

It is clear that the 1-D CosForce model closely resembles the FVD model \citep{jiang2001full}, also the 2-D version \citep{lv2013two}. The difference is that the proportional coefficient does not undergo the same scale transformation as the velocity difference. The main issue lies in the inconsistency properties of speed between different backgrounds: in traffic, a small speed difference is indicative of a stable flow state, whereas in crowds, a small speed difference may be interpreted as mutually deceleration due to conflicts.

\textbullet\ \textbf{2-D OV model, when \(\mathbf{\alpha} = \mathbf{0}\)}

When $\alpha = 0$, the 1-D dynamical equation reduces to Eq.\ref{16}, which exhibits high similarity to the OV model \citep{bando1995dynamical}, as well as to the 2-D OV model \citep{nakayama2005instability}.

\begin{equation}\label{16}
f = m_i a = \frac{m_i}{\tau} \left( \max \left( \min \left( \frac{ {d}_{ij}  - r_{ij}}{t_h},  {v}_{\max}  \right), 0 \right) - {v_i} \right)
\end{equation}

\section{Numerical Validation} \label{section3}

In this section, the properties of the CosForce model are examined based on empirical results. The investigated scenarios include single-file motion and unidirectional flow.

\subsection{Single-file Dynamics} \label{subsection3.1}

\begin{figure}[ht!]
\centering
\includegraphics[scale=0.45]{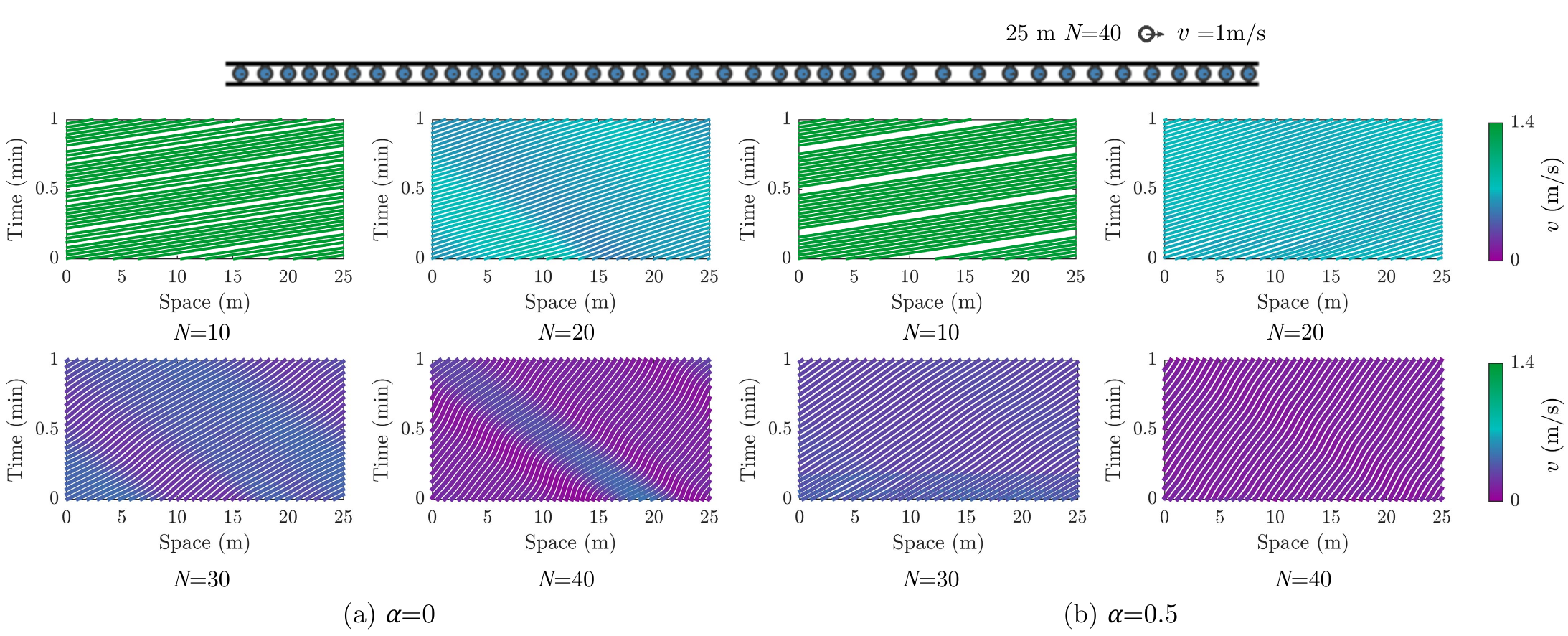}
\caption{Time-space diagram at different densities: (a) \(\alpha = 0\), (b) \(\alpha = 0.5\).}
\label{fig6}
\end{figure}

First, we simulated the single-file motion under periodic boundary conditions, with the variable parameter \(\phi\) set to \(\pi/3\), as illustrated in Fig.\ref{fig6}. To compare the differences in model performance between \(\alpha = 0\) and \(\alpha = 0.5\), we set \(N = \) \(10\), \(20\), \(30\) and \(40\) in the simulation to observe the dynamics of single-file pedestrians under varying densities. The pre-simulation period was set to 500 steps, and simulation data for time steps 501-2300 were analyzed after the pre-simulation (corresponding to 60 s). Since random noise was excluded, the simulation results were deterministic based on the initial conditions. The results for \(\alpha = 0\) and \(\alpha = 0.5\) are shown in Fig.\ref{fig6}. From the perspective of stability, when \(\alpha = 0.5\), the model demonstrated superior performance. During the observation period, the traffic flow exhibited synchronization characteristics and no stop-and-go waves were observed at \(N = 40\). This stability property aligns with the advantages of the FVD model over the OV model.

\subsection{Unidirectional Flow} \label{subsection3.2}

\begin{figure}[ht!]
\centering
\includegraphics[scale=0.48]{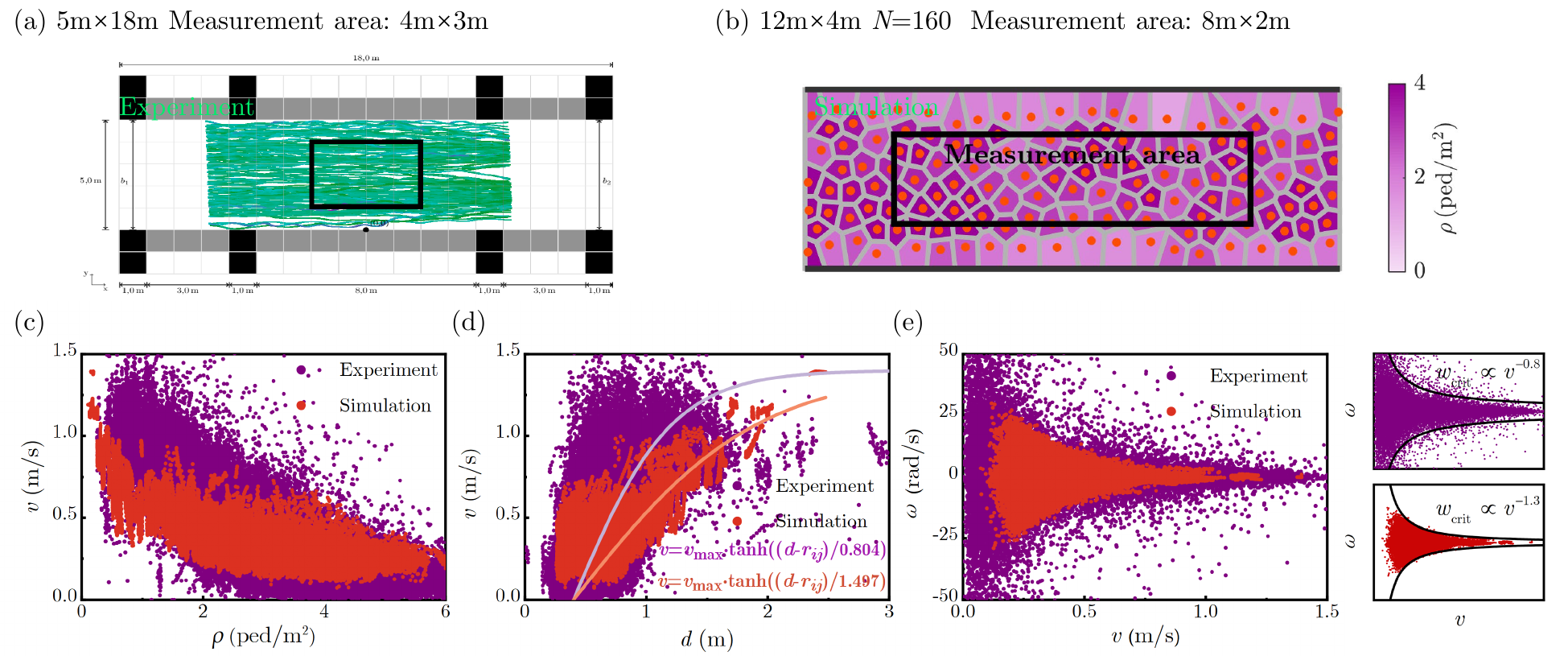}
\caption{Experimental and simulation comparison of unidirectional flow dynamics. (a) Experimental scenario. (b) Simulation scenario. (c) Relationship between local density and speed. (d) Relationship between headway and speed. According to the simulation settings, the parameters \(v_{max}\) and \(r_{ij}\) in the fitting function were set to 1.4 m/s and 0.4 m, respectively. (e) Relationship between angular velocity and speed.}
\label{fig7}
\end{figure}

We conducted simulations of pedestrian unidirectional flow scenarios in a channel with periodic boundary conditions, as shown in Fig.\ref{fig7}. In the simulations, the variable parameters set as \(\phi=\pi/3\) and \(\alpha=0.5\). We increased the number of pedestrians incrementally by 10 per simulation, running a total of 16 simulations to cover scenarios from \( N = 10 \) to \( N = 160 \). The pre-simulation duration was 500 time steps. Subsequently, data were acquired at 30 Hz for 100 time steps (corresponding to 3.33 s). Fig.\ref{fig7} illustrates the fundamental relationships of unidirectional pedestrian flow, including density versus speed, headway versus speed, and angular velocity versus speed. Due to the fixed sampling interval, data obtained from high-density simulations are significantly more abundant than from low-density conditions, resulting in an uneven data distribution. The empirical analysis data are derived from trajectory observations in the unidirectional experiment \citep{cao2017fundamental}, with detailed experimental configuration information provided in Appx.\ref{Appendix A}.

\textbf{Local density versus Speed}  

Fig.\ref{fig7}(c) presents the relationship between pedestrian speed and local density, which conforms to the characteristics of the empirical speed-density curve, showing an inverse proportional relationship in constrained conditions.

\textbf{Headway versus Speed} 

Fig.\ref{fig7}(d) shows the relationship between pedestrian speed and the relative distance to the nearest neighbor. Specifically, based on the constant time headway property, speed exhibits a linear relationship with the nearest neighbor distance under constrained conditions. The variation trend aligns with empirical findings. A comparison of the data fitting results between experiments and simulations reveals that the experimental time headway (0.804 s) is lower than the simulated result (1.497 s), with the fitting function given by Eq.~\ref{17}. This discrepancy stems from parameter settings, as the simulation uses a fixed \(t_h\) of 1.3 s throughout the study.

\begin{equation}\label{17}
v = {v_{\max }} \cdot \tanh \left( {\frac{{d - {r_{ij}}}}{{{t_h}}}} \right)
\end{equation}

\textbf{Angular velocity versus Speed}

Fig.\ref{fig7}(e) shows the relationship between pedestrian angular velocity and speed, with the constraint exponent of maximum angular velocity corresponding to speed approximately equal to -1.3, slightly lower than our empirical observation (where the constraint exponent is approximately -0.8 \citep{wang2025kinematic}).

\section{Simulation of "Phase separation" in crowds} \label{section4}

Walking and interacting with others are activities that everyone engages in daily. Therefore, preserving intuition is a crucial aspect when investigating crowd dynamics. With this consideration, we have focused on observing fundamental parameters in statistics, excluding any composite parameters. In crowd dynamics, "phase separation" refers to the phenomenon in which separation occurs between groups with different directions or states, induced by spontaneous order. Common examples include lane formation, stripe formation, and cross-channel formation \citep{VideoExample1}, among others. However, effective metrics for quantitatively estimating self-organization phenomena in crowds are still limited. Based on the mechanisms of self-organization (entropy decrease and steady state), the average normalized speed \(\langle v \rangle\), the order parameter \(\Phi\), and the average normalized velocity \(\langle\mathbf{v}\rangle\) were defined, as given in Eq.\ref{18}. These metrics serve as quantitative benchmarks for evaluating three distinct aspects of crowd: individual efficiency, system entropy (negatively correlated with the order parameter in crowds), and collective motion at the crowd scale. In the simulation of pedestrian self-organization, since the desired direction is predefined, the order parameters for pedestrians with different directions were calculated independently and averaged to avoid mutual cancellation.

\begin{equation}\label{18}
\left\{
\begin{aligned}
\langle v \rangle &= \frac{1}{Nv_{max}} \sum\limits_{i \in N} \lVert\mathbf{v}_i \rVert 
&& \text{(average normalized speed)} \\
\Phi &= \frac{1}{N} \left\|  \sum_{i=1}^{N} \frac{\mathbf{v}_i}{\lVert\mathbf{v}_i \rVert} \right\|
&& \text{(order parameter)} \\
\langle \mathbf{v} \rangle &= \frac{1}{Nv_{max}}  \sum_{i \in N} \mathbf{v}_i  
&& \text{(average normalized velocity)}
\end{aligned}
\right.
\end{equation}

The simulation geometry is set to 8 m × 8 m with fully periodic boundary conditions. In the initial state, all pedestrians are assigned an initial speed of 0 and distributed randomly. The simulation duration is 3000 time steps (corresponding to 100 s).

\subsection{Lane formation} \label{subsection4.1}

\begin{figure}[ht!]
\centering
\includegraphics[scale=0.6]{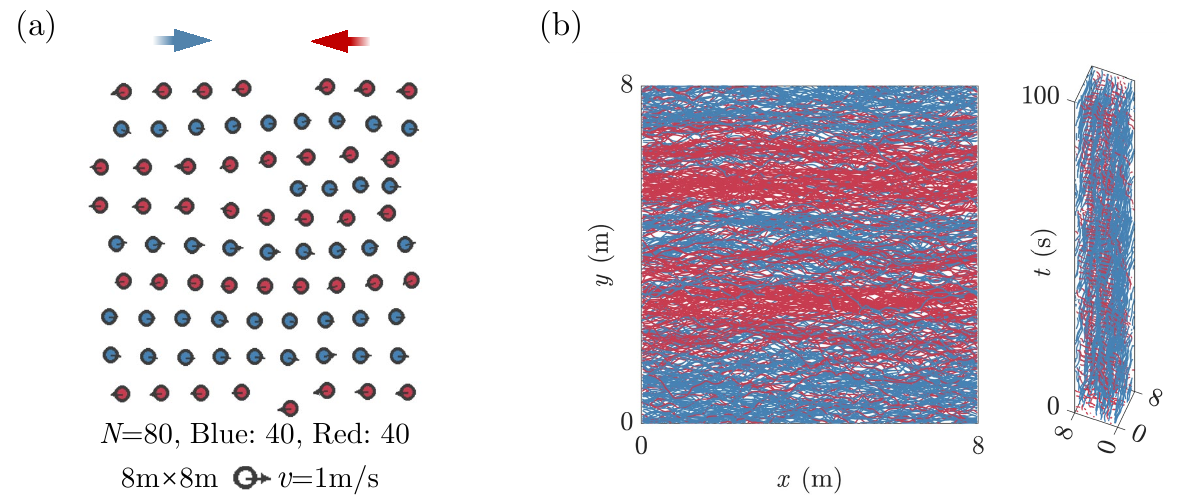}
\caption{Illustration of the lane formation phenomenon. (a) Simulation snapshot. (b) Pedestrian trajectory patterns during the simulation.}
\label{fig8}
\end{figure}

\begin{figure}[ht!]
\centering
\includegraphics[scale=0.7]{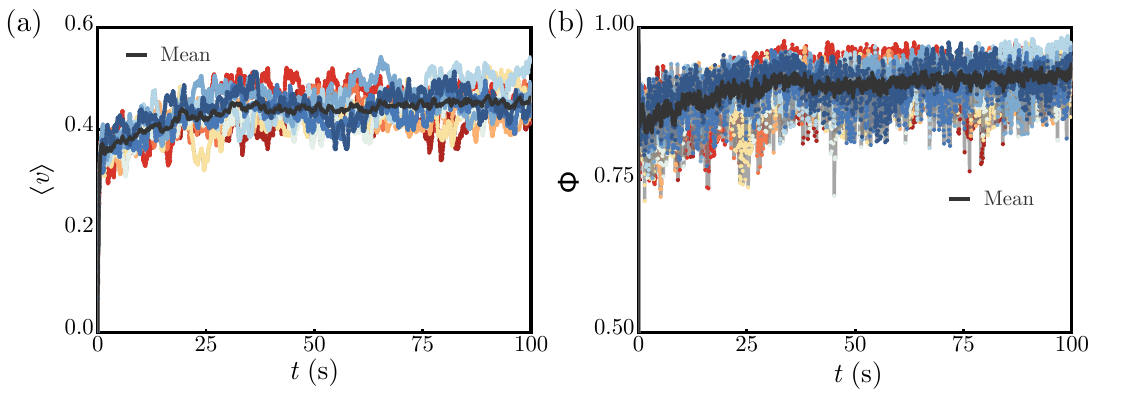}
\caption{Time series variations of the average normalized speed (a) and the order parameter (b) in the simulation of lane formation phenomenon. The order parameters for different flow directions were calculated independently.}
\label{fig9}
\end{figure}

First, we conducted simulations of lane formation phenomena in bidirectional pedestrian flows. The total number of pedestrians was set to 80, with a directional distribution of 40 : 40. The variable parameters were configured as \(\phi = \pi/2\), and \(\alpha = 0.5\). Fig.\ref{fig8} displays a snapshot of the self-organization phenomenon observed in the simulation, along with the time series evolution of the corresponding trajectories. We performed 10 independent simulations, and the trends of the average normalized speed and the order parameter are shown in Fig.\ref{fig9}. The simulation results indicate that, during a short initial period (approximately 30 s), the average normalized speed increases and stabilized. Similarly, the order parameter exhibits an increase before reaching a steady state. These results demonstrate that, from the perspectives of efficiency and entropy, the simulated crowd system achieves a stable state. Thus, the self-organization process can be quantitatively evaluated.

\subsection{Stripe formation} \label{subsection4.2}

\begin{figure}[ht!]
\centering
\includegraphics[scale=0.6]{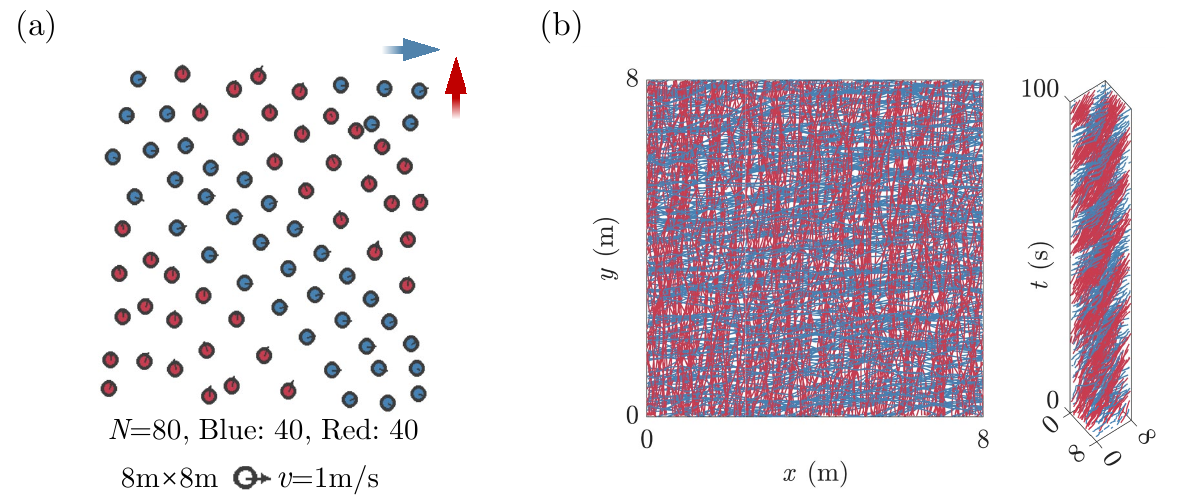}
\caption{Illustration of the stripe formation phenomenon. (a) Simulation snapshot. (b) Pedestrian trajectory patterns during the simulation. The order parameters for different flow directions were calculated independently.}
\label{fig10}
\end{figure}

\begin{figure}[ht!]
\centering
\includegraphics[scale=0.7]{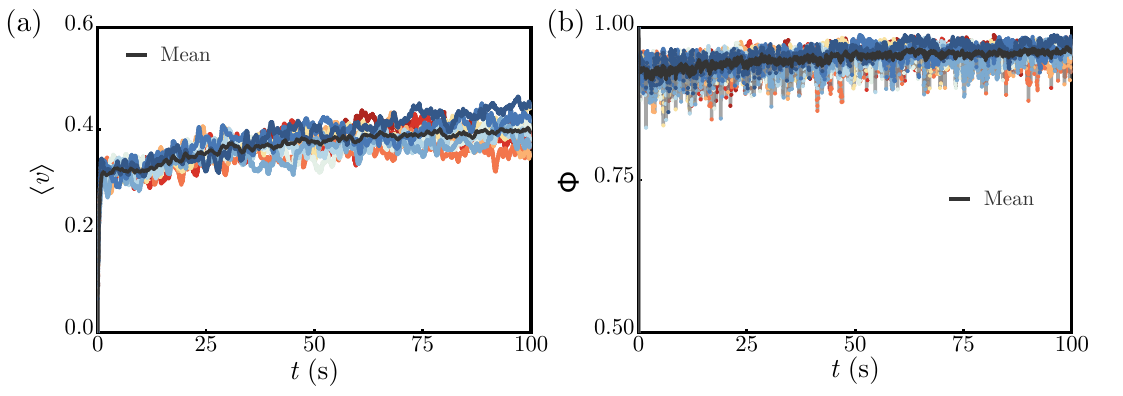}
\caption{Time series variations of the average normalized speed (a) and the order parameter (b) in the simulation of stripe formation phenomenon.}
\label{fig11}
\end{figure}

Additionally, we conducted a simulation of the phenomenon of  stripe formation in cross flow. The total number of pedestrians was set to 80, with a directional distribution of 40 : 40. The variable parameters were configured as \( \phi = \pi/3 \), and \( \alpha = 0.5 \). Fig.\ref{fig10} shows a snapshot of the self-organization phenomenon observed in the simulation, along with the time series evolution of the corresponding trajectories. Fig.\ref{fig10} clearly reveals the phase separation process between pedestrians moving in opposite directions. Ten independent simulations were performed, and the trends of the average normalized speed and order parameter are presented in Fig.\ref{fig11}. The variation is similar to that observed in Fig.\ref{fig10}, as the average normalized speed and order parameter increase, the system transitions into a stable phase. In terms of mean value, this trend remains consistently steady. These characteristics indicate the spontaneous formation of order within the crowd, which typically occurs within a time frame ranging from several seconds to a few tens of seconds.

\subsection{Cross-channel formation} 
\label{subsection4.3}

The final scenario in the simulation of phase separation is referred to as "cross-channel formation", which describes the phenomenon where pedestrians, when crossing through static crowds, form a series of stable cross-channels. A field video of this phenomenon can be observed in a train station \citep{VideoExample1}. In our simulations, the total number of pedestrians was set as 250, with a proportionally varied pedestrian state configuration (red refer dynamic pedestrian and blue refer static pedestrian). The variable parameters were set as \( v_{\rm{max}} = 1.4 \; \text{m/s} \) (dynamic) or \( 0 \; \text{m/s} \) (static), \( \phi = \pi/2 \) (dynamic) or \( \pi \) (static), and \( \alpha = 0.5 \).

\begin{figure}[ht!]
\centering
\includegraphics[scale=0.6]{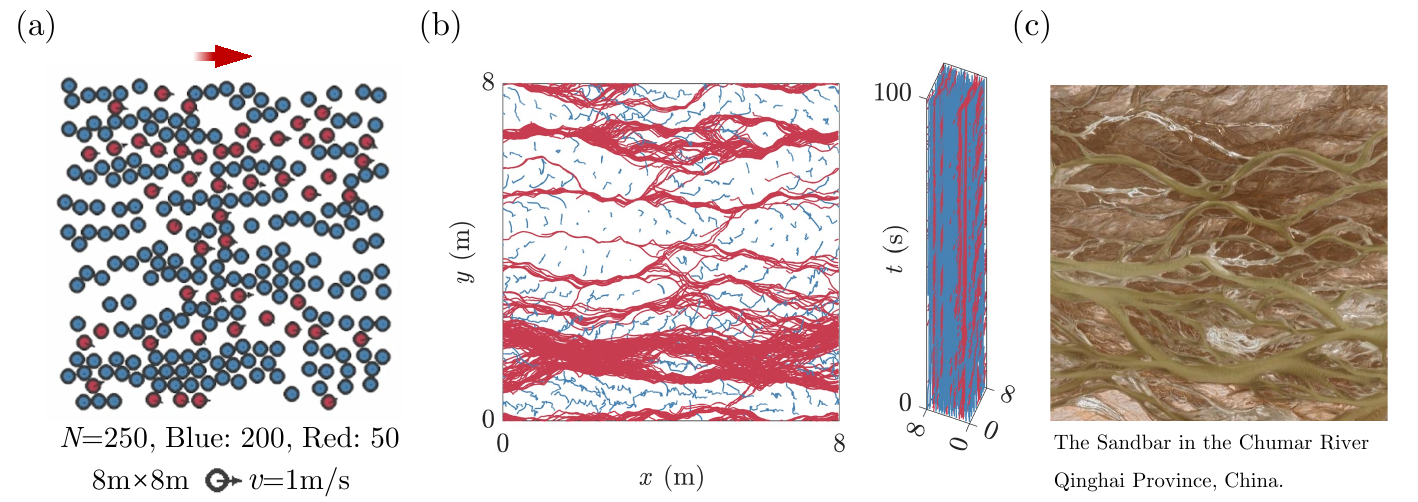}
\caption{Illustration of the cross-channel formation phenomenon. (a) Simulation snapshot. (b) Pedestrian trajectory patterns during the simulation. (c) Formation of sandbars induced by solid-liquid flow interaction. Source: Google Earth, location: 35$^\circ$13'07"N 93$^\circ$55'09"E.}
\label{fig12}
\end{figure}

\begin{figure}[ht!]
\centering
\includegraphics[scale=0.7]{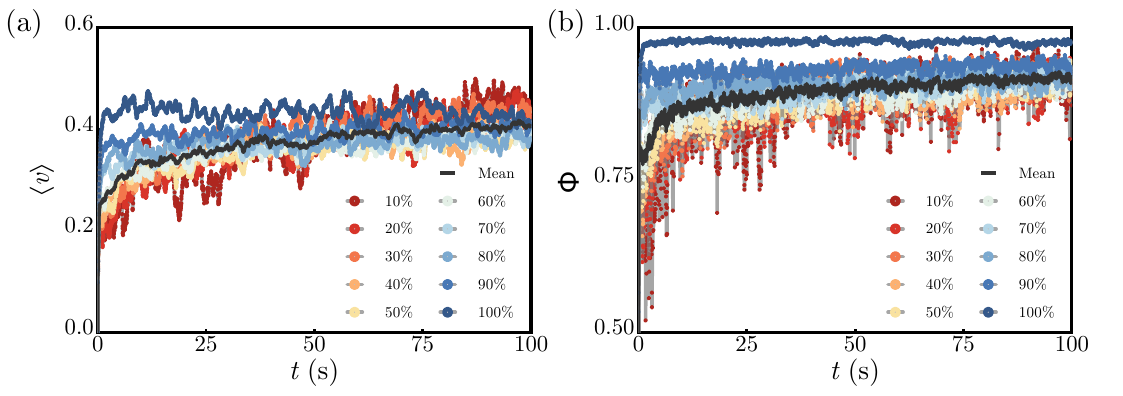}
\caption{Time series variations of the average normalized speed (a) and the order parameter (b) of dynamic pedestrians in the simulation of the cross-channel formation phenomenon. The mean values were calculated based on data from simulations with the ratio of dynamic pedestrians ranging from 10\% to 90\%. The simulation with a 100\% ratio of dynamic pedestrians was treated as a control group and excluded from the statistical analysis.}
\label{fig13}
\end{figure}

Fig.\ref{fig12} illustrates the snapshot of self-organizing phenomena observed in the simulation, along with the time series evolution of the corresponding trajectories. From the snapshot, the formation of distinct cross channels can be clearly observed. This mechanism bears a strong resemblance to the evolution of sandbars, as discussed in the context of liquid-solid flow interactions. Ten independent simulations were conducted, each corresponding to a different ratio of dynamic pedestrians. Since static pedestrians remain stationary in most scenarios, only dynamic pedestrians were considered to avoid the dilution effect in statistics. The variation trends of the average normalized speed and the order parameter are shown in Fig.\ref{fig13}. The statistical mean is calculated based on the average data for configurations ranging from 10\% to 90\%, with the 100\% ratio serving as the control group representing unidirectional flow. As seen in Fig.\ref{fig13}, similar self-organization trends can be observed. Specifically, both the average normalized speed and the order parameter show a steady increase before eventually stabilizing. Clearly, as an evolutionary process similar with liquid-solid flow interactions, the phenomenon of cross-channel formation exhibits clearly characteristics of self-organization.

\section{Simulation of mass gathering} \label{section5}

Self-organization phenomena demonstrate the order and vitality exhibited by a crowd as an organic system. However, this is not the entirety of crowd dynamics. In some cases, crowds may exhibit characteristics entirely opposite where their motion become highly chaotic and locally coherent mechanisms emerge (velocity correlation, induced by pushing and shoving), leading to destructive outcomes. As a open challenge, the simulation of crowd collapse is regarded as a critically important objective of modeling efforts. Specifically, it aims to apply models to simulate crowd accidents and uncover insights that are difficult to obtain through empirical investigations. This represents a critical aspect for understanding the underlying dynamics of crowds and managing them effectively, though it may also be the most challenging aspect. 

The continuous dynamism of crowds is one of the fundamental causes of crowd crush. Some characteristics of crowd oscillation have been empirically observed, with contributions from pioneering works \citep{gu2024emergence, ma2013new, echeverria2022spontaneous} on the phenomenon of crowd quakes. Before proceeding with the simulation and analysis, it is essential to delineate the limitations of the simulations presented in this paper. As a decentralized system closely intertwined with our daily lives, the dynamics exhibited by crowds are characterized by diversity, stemming from their self-propelled property and social nature. A most intuitive observation is that the dynamics of crowds vary significantly under different circumstances. This variation is associated with the scale of the crowd, the environment, and the intensity of self-propelled behavior induced by social objectives. Among these, the truly dangerous situations arise in dynamical dense crowds, which have prompted lots quantitative analyses based on the properties of velocity fields \citep{feliciani2018measurement,zanlungo2023pure}. The crowd system simulated in this paper focuses on the crowd with dynamism property, while static crowds are not considered in this discussion.
Moreover, the two-dimensional simulation in this study cannot capture mechanisms such as pedestrian falls, which may lead to distortions in simulating crowd collapse. Therefore, awareness of the model’s limitations is essential. Our analysis will commence with a specific phenomenon  within crowds, namely the "catfish effect," to investigate the impact of dynamic pedestrians on the system of crowd context. Subsequently, we will discuss more generalized crowd dynamics and explore the critical transition processes in dense crowds.

\subsection{"Catfish effect" in Crowd} \label{subsection5.1}

The "catfish effect" in crowd context can be defined as the increase in the overall dynamism of the crowd induced by a minority of dynamic pedestrians, quantitatively marked by the system's speed gain exceeding the speed increment of dynamic pedestrians. When dynamic pedestrians enter the crowd, adjacent static individuals will gradually adjust their velocity, and the random motion becomes more organized (i.e., polarized). The symmetry breaking resulted in a significant increase in the locally speed.  At the microscopic level, this mechanism can be understood as the local polarization of crossing individuals, which has been discussed in lots experiments and models \citep{bonnemain2023pedestrians, nicolas2019mechanical}. This paper focuses on the impact of dynamic pedestrians within the crowd scale and explores the effect of symmetry variation.

\begin{figure}[ht!]
\centering
\includegraphics[scale=0.5]{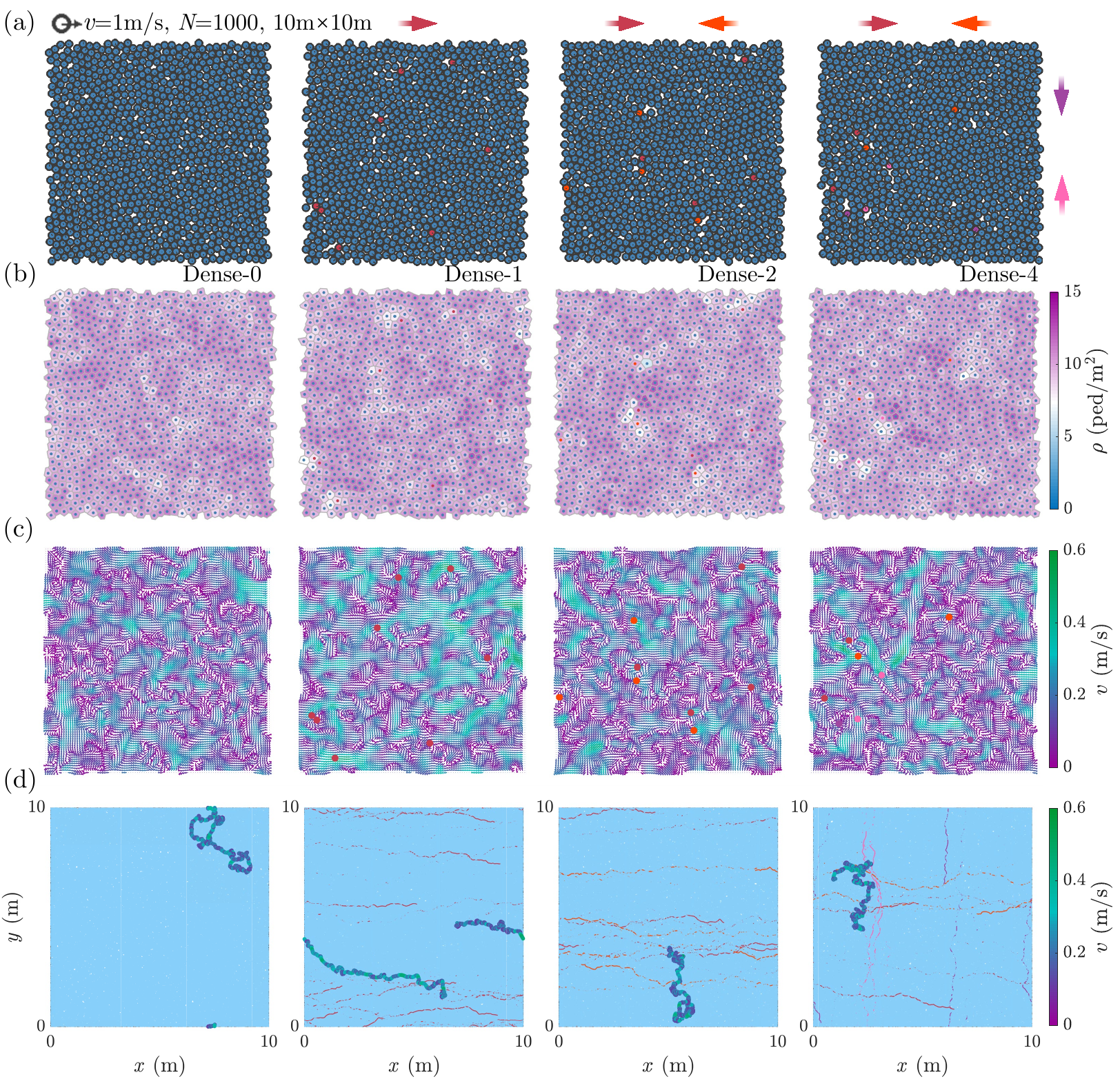}
\caption{Simulation of the "catfish effect" phenomenon in dense crowds. (a) Snapshots of the simulation at 50s: arrows and colors indicate the direction of dynamic pedestrians. (b) The local density distribution at 50 s, with colors representing different types of pedestrians. (c) The velocity field distribution at 50 s, where dots denote the positions of dynamic pedestrians. (d) Trajectories over the 100s simulation period, with bold trajectories highlighting the motion of randomly selected static pedestrians.}
\label{fig14}
\end{figure}

To simulate these mechanisms, variable parameters were set as \({v}_{max} = 1.4 \; \text{m/s}\) (dynamic) or \( 0.6 \; \text{m/s} \; [0,0]\) (static), \( \phi = \pi/2 \) (dynamic) or \( \pi \) (static), and \( \alpha = 0.5 \). The zero vector \([0, 0]\) represents the direction of the maximum velocity. According to Eqs.\ref{11} and \ref{12}, under these parameter settings, the self-driven force will try to slow down static pedestrians to a speed of 0, while the repulsive force will be induced by nearest neighbor, scale corresponds to a maximum speed of 0.6 m/s. This setup leads to static pedestrians being highly sensitive to spatial variations and leads to a structured distribution, which is characteristic of dynamic crowds. Overall, the crowd mainly consists of static pedestrians, providing the background environment of the system. Our aim is to study changes in the state caused by dynamic pedestrians and explore the potential phenomenon of the "catfish effect", which originates from our earlier conjecture \citep{wang2023exploring}. The simulation space is set as an 10 m $\times$ 10 m with fully periodic boundary conditions. The simulated crowd size is set to 1000, corresponding to a simulation space of 100 m\textsuperscript{2}, with a global density of 10 ped/m\textsuperscript{2}. At such a density, pedestrians are tightly packed together, and as a consequence, the interactions within the crowd are primarily governed by contact forces. In the initial state, all pedestrians are randomly distributed, with an initial speed of 0 m/s. The simulation time for a single scenario is 3000 time steps (corresponding to 100 s).

A simulation scenario without dynamic pedestrians served as the control group for comparative analysis against three intrusion types: unidirectional intrusion (8 dynamic pedestrians), bidirectional intrusion (8 dynamic pedestrians, configured as 4:4), and cross-directional intrusion (8 dynamic pedestrians, configured as 2:2:2:2). Based on the directional configuration, the simulations are labeled as Dense-0, Dense-1, Dense-2, and Dense-4, respectively. Fig.\ref{fig14}(a) presents snapshots of the four simulation groups at the moment of 50 s. Figs.\ref{fig14}(b) and (c) show the corresponding distributions of the local density field and velocity field. In the simulation results, compared to the control group (Dense-0), the introduction of dynamic pedestrians in Dense-1 led to a significant increase in speed, while the speed changes in Dense-2 and Dense-4 were less pronounced. The distribution patterns of the trajectories further support this observation, as illustrated in Fig.\ref{fig14}(d).

\begin{figure}[ht!]
\centering
\includegraphics[scale=0.7]{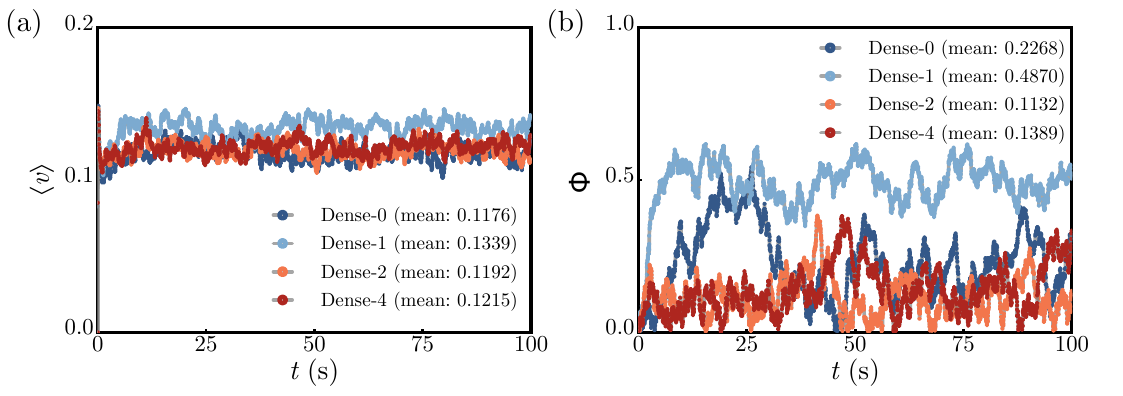}
\caption{Time series variations of the mean normalized speed (a) and the order parameter (b) across different crowd configurations.}
\label{fig15}
\end{figure}

\begin{figure}[ht!]
\centering
\includegraphics[scale=0.6]{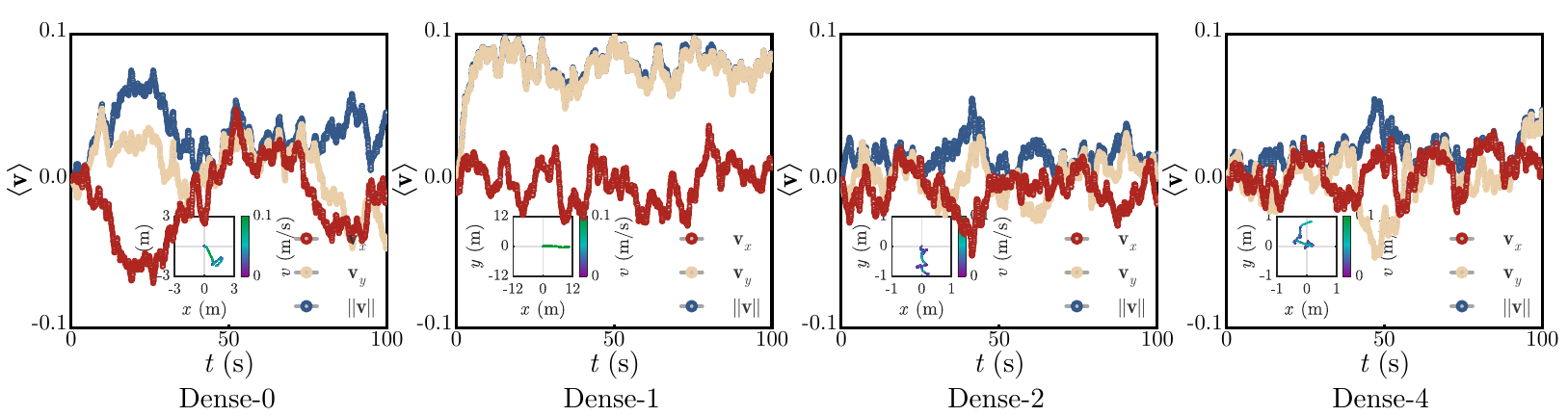}
\caption{Time series variations of the average normalized velocity across different crowd configurations. The trajectories in the subplots illustrate the motion pattern in the crowd scale.}
\label{fig16}
\end{figure}

Fig. \ref{fig15} shows the time series variations of the mean normalized speed and the order parameter across different crowd configurations. In the Dense-1 simulation, the speed gain reaches approximately 16.3 units, more than twice the maximum possible increment (less than 8 units). However, the underlying mechanism remains unclear in the Dense-2 and Dense-4 simulations. The order parameter data suggest that the polarization tendency in Dense-1 remains the strongest, while Dense-2 and Dense-4 show very similar trends, both falling below the control group (Dense-0). Fig. \ref{fig16} presents the variation trends of the average normalized velocity under different crowd configurations. It can be observed that in the unbalanced dynamic pedestrian configuration (Dense-1), the crowd exhibits a stronger polarization tendency. In contrast, this difference is not evident in the balanced dynamic configurations. This effect may be attributed to global statistical observations masking the polarization tendency in relative orientations.

\subsection{Crowd collapse: from crowd quake to crowd avalanche} \label{subsection5.2}

In Section \ref{subsection5.1}, beyond analyzing the potential "catfish effect" within the crowd, we identified variations in the order parameter, indicating that changes in collective symmetry may be related to the onset of crowd collapse. To simulate crowd collapse and investigate the underlying mechanism, we established a simulation scenario where the global density increases incrementally over time. The settings and parameters, consistent with the simulation in section \ref{subsection5.1}, consist of only static pedestrians for generality. In the simulation, the initial crowd density is 5 ped/m\textsuperscript{2}. To create a continuously increasing density scenario for time series observation, we add one pedestrian to the system every 30 time steps. The total simulation duration is 18,000 time steps (corresponding to 600 seconds), during which the maximum density will reaches 11 ped/m\textsuperscript{2}, representing the theoretical upper limit of crowd density.

\begin{figure}[ht!]
\centering
\includegraphics[scale=0.56]{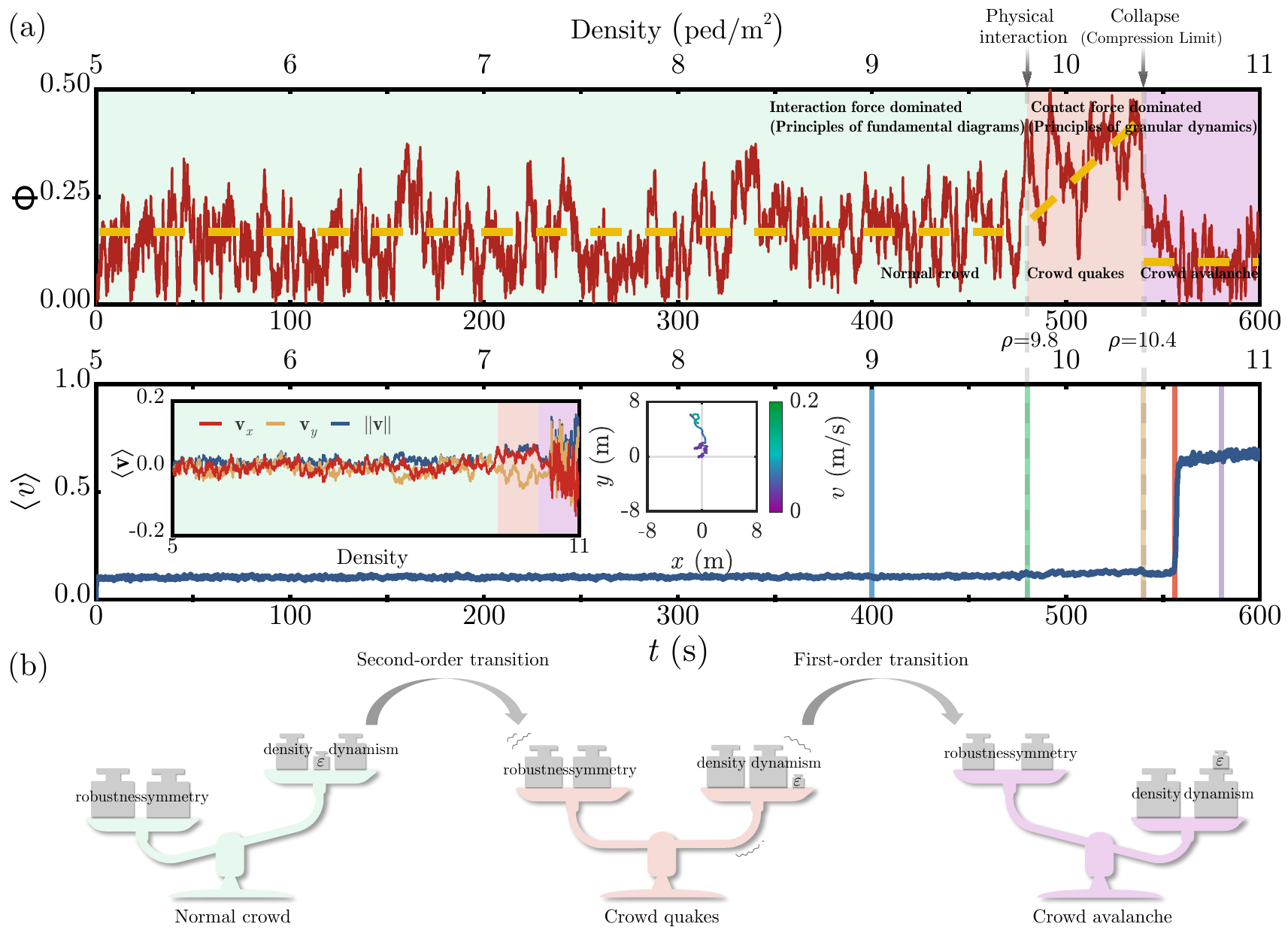}
\caption{Critical transition process of crowd system. (a) time series variation of the order parameter and average normalized speed, (b) balance scale illustration of the critical transition in crowd dynamics. The terms "crowd quakes" and "crowd avalanche" were adopted from \citet{ma2013new} and \citet{feliciani2022introduction}, \(\varepsilon\) denotes the perturbation.}
\label{fig17}
\end{figure}

The time series curves of the order parameter and average normalized speed are shown in Fig.\ref{fig17}(a). From the trend of the order parameter, the state transition of the crowd can be identified. The first transition occurs near \(\rho = 9.8 \, \text{ped/m}^2\). When the pedestrian density exceeds this threshold, the order parameter begins to rise from a state of steady fluctuations. This transition shows a second-order transition, where the space between pedestrians is compressed to a delicate equilibrium point. The motion of the crowd shifts from a self-propelled process (constrained by the fundamental diagram) to the process of granular dynamics (action equals reaction). The second transition occurs near \(\rho = 10.4 \, \text{ped/m}^2\), where the order parameter undergoes a sudden transition, indicating a first-order transition. At this moment, the crowd transitions from a highly compressed state to a collapsed state exhibiting chaotic characteristics. Transition of average normalized speed shows a delay, which occurs near \(\rho = 10.56 \, \text{ped/m}^2\), approximately 16 s later than the transition in the order parameter. The subplot in Fig.\ref{fig17}(a) shows the temporal trend of the average normalized velocity, revealing that the motion of the crowd also undergoes similar transitions. The trajectory plot shows the crowd's collective motion in the three states: low polarization \(\rightarrow\)  high polarization \(\rightarrow\)  low polarization with high dynamism.

The critical transition process can be intuitively explained through the analogy of a tipping balance scale, as shown in Fig.\ref{fig17}(b). At low densities, the crowd system exhibits adaptive behavior. Its state remains largely unaffected by environmental influences. As both density and dynamism increase, the crowd tends to a more compact and structured distribution. This is accompanied by a decline in robustness and tendency of polarization. Under such conditions, the crowd exists in a delicate equilibrium, analogous to a balanced scale, where even slight perturbations can induce a transition. The transition from equilibrium to deflection is similar to the propagation of local imbalance among individuals in the crowd. This imbalance gradually spreads and evolves into a domino effect. Such a process accounts for the delayed onset of speed transition observed in the simulation.

\begin{figure}[ht!]
\centering
\includegraphics[scale=0.53]{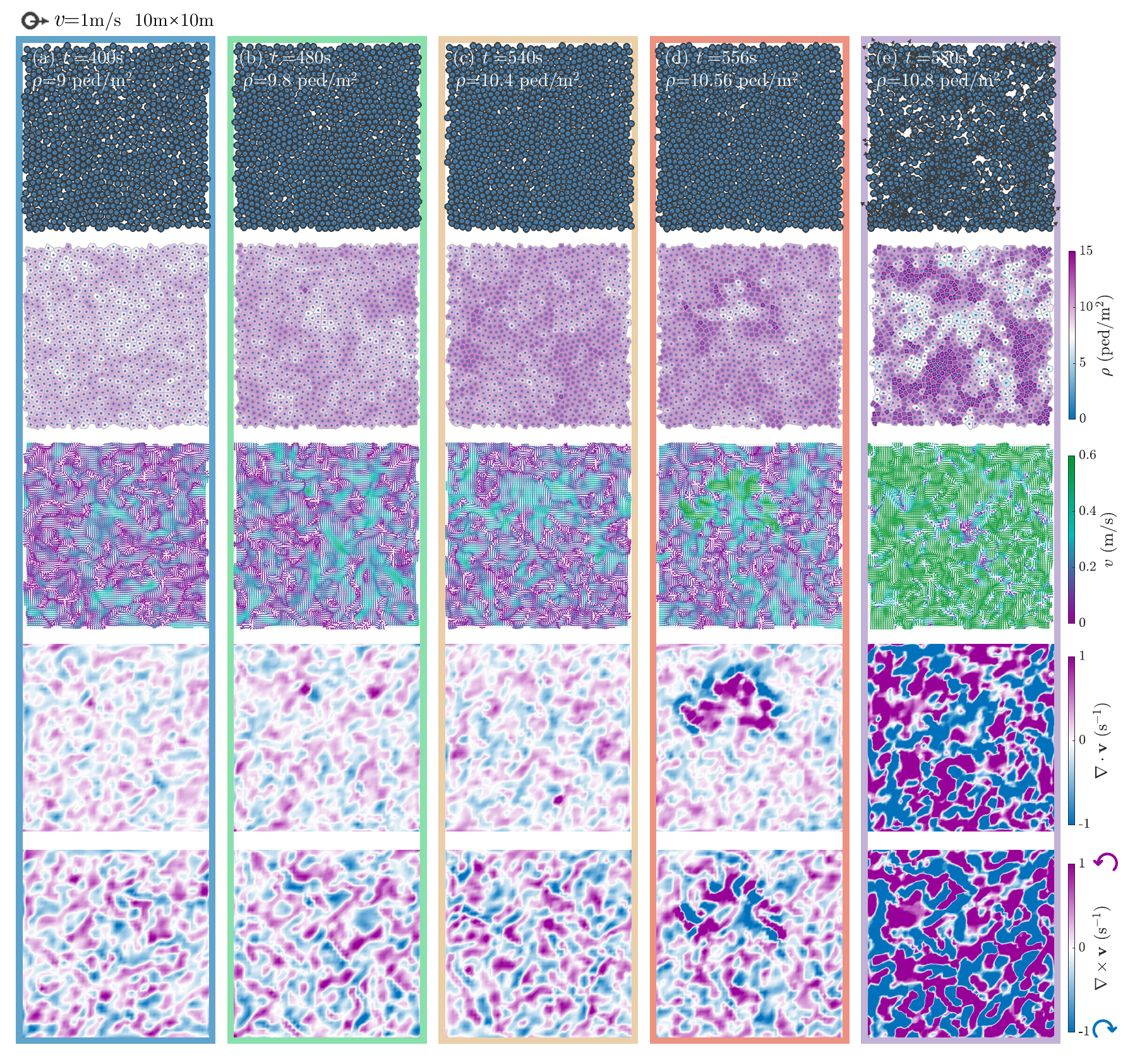}
\caption{Simulation results for crowd densities of 9, 9.8, 10.4, 10.56, and 10.8 ped/m\(^2\). From top to bottom, the simulation snapshots, local density field, velocity field, divergence, and vorticity distributions are presented respectively. Based on the color index, the crowd can be localized to the time
 series information shown in Fig.\ref{fig17}.}
\label{fig18}
\end{figure}

Critical moments observed during the simulation correspond to the times marked by vertical lines in Fig.\ref{fig17}(a), with densities of 9, 9.8, 10.4, 10.56, and 10.8 ped/m\(^2\), respectively. We present the simulation scenarios along with the corresponding density fields, velocity fields, and the field of divergence and vorticity, as shown in Fig.\ref{fig18}. At the moment corresponding to \(\rho = 9\)~ped/m\(^2\), the crowd remains in a stable state. Pedestrians exhibit a relatively uniform spatial distribution. When the density increases to 9.8~ped/m\(^2\), spatial contraction among individuals and an increase in pedestrian speed can be observed. This corresponds to the moment of the second-order transition in Fig.~\ref{fig17}. As the density further rises to 10.4~ped/m\(^2\), the system reaches the first-order transition point. At this stage, pedestrian compression approaches a critical limit, beyond which further compression becomes nearly impossible. Mutual physical forces and motion interactions render individuals almost unable to maintain balance. This imbalance initially occurs at the individual level and indicates a disruption of force equilibrium. In real-world scenarios, such a condition may lead to pedestrian falls. This imbalance breaks the stability of the crowd system and marks the tipping point when the crowd begins to shift. The moment \(\rho = 10.56\) ped/m\(^2\) corresponds to the process of crowd collapse, as shown in Fig.\ref{fig18}(d), where a domino-like propagation mechanism is clearly visible. The immense internal energy generated by crowd compression erupts upon reaching the compression limit, resulting in the occurrence of crowd collapse. Fig.\ref{fig18}(e) demonstrates the crowd collapse phenomenon observed in the simulation. The periodic boundary conditions and the exclusive consideration of force transmission overlook potential falling processes during crowd collapse. This leads to a notable discrepancy from real scenarios.

\subsection{Discussion} \label{subsection5.3}

\begin{figure}[ht!]
\centering
\includegraphics[scale=0.6]{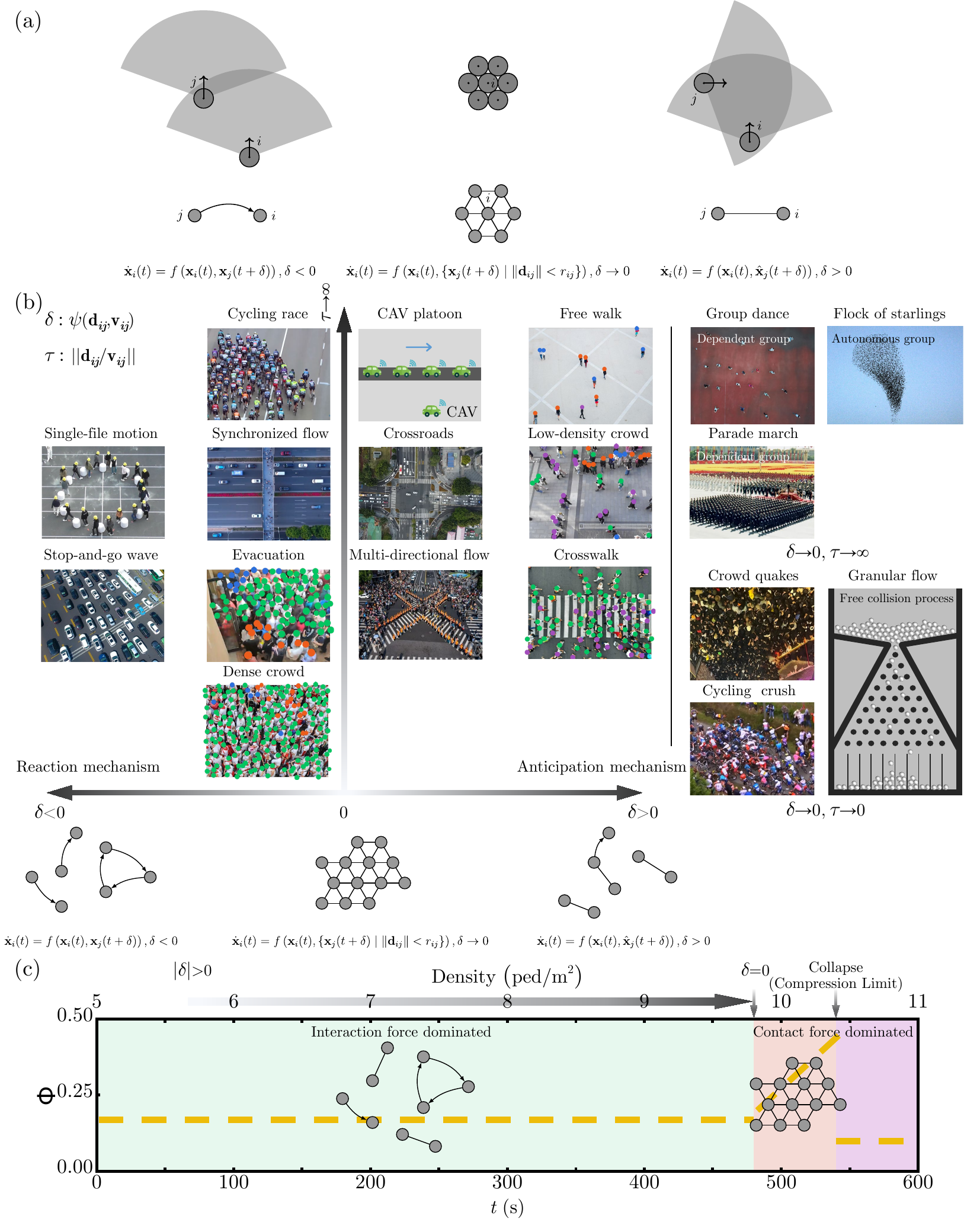}
\caption{Interaction graph of crowd: the topology and corresponding efficiency of the network govern the collective properties of the crowd. In subfigure (b), $\psi$ denotes the phase difference function. For further explanation of the space-speed time delay ($\delta$), please refer to \citet{wang2024cosin}.}
\label{fig19}
\end{figure}

In sections \ref{subsection5.1} and \ref{subsection5.2}, we discussed the "catfish effect" and the process of crowd collapse, and provided an explanation based on the processes of critical transition. From the simulation results, we identified several densities of
tipping point for crowd transitions, specifically $\rho =$ 9.8 and 10.4 ped/m\(^2\). These values are applicable only within the context of this simulation. Human body size varies significantly by gender, age, and other physiological factors, density value in the context of pedestrian crowd is an insufficient metric. More importantly, it is essential to understand the critical conditions under which individuals lose control of their motion due to contact. All insights in section \ref{section5} were derived retrospectively from the perspective of simulation with inherently limitations and potential errors. Based on the presented results, several issues need to be discussed in this section.

\textbullet\ \textbf{Network dynamics perspective on crowd critical transition.}

The network mapping of crowd interactions facilitates the analysis of critical transition processes from the perspective of network dynamics, thereby enabling a more concise understanding. As illustrated in Fig. \ref{fig19}(a), in non-contact scenarios, the anticipation and reaction behaviors of pedestrians can be characterized using mappings of arc connections (feedback mechanism) and edge connections (feedforward mechanism). The interactions among dense crowds correspond to structured networks. Such relationships can be extended to broader contexts and incorporated into a unified framework encompassing diverse collective systems, as demonstrated in Fig. \ref{fig19}(b). The topological structure of interaction graphs governs the dynamical evolution of crowds, where the connectivity influences the system's response to conditional variations. In networks with incomplete connectivity that result in modular structures (characterized by low crowd density where pedestrians maintain certain distances), the system often exhibits a gradual adaptive capacity to changes. In contrast, within structured networks (featuring high crowd density with direct physical contact among pedestrians), individual disturbances are amplified through synergistic effects among linked nodes, ultimately driving the crowd system toward abrupt collapse under critical perturbations (critical transition) \citep{scheffer2012anticipating}.

As illustrated in Fig. \ref{fig19}(c), in our simulation, the network undergoes a gradual transition from a sparse configuration to a structured architecture as density increases. During this process, the $ \delta $ between nodes progressively converges to zero, leading to enhanced response efficiency in the crowd system. Within crowd contexts, this represents a dangerous property. When the $ \delta \to 0 $, it indicates the onset of mutual compression within the crowd, manifested through increased velocity coherence and correlation length. Should this trend continue to develop, the occurrence of a critical transition at the tipping point becomes inevitable.

\textbullet\ \textbf{Whether dynamic (or aggressive) pedestrians would affect the process of crowd collapse.}

Potential intentional pushing was reported during the Itaewon crowd crush incident \citep{kim2022crush}. Similar aggressive behaviors, as manifestations of crowd heterogeneity, require attention in simulations. In our simulation, introduction of dynamic pedestrians significantly destabilizes the state of the crowd. From the perspectives of mobility (average normalized speed) and symmetry (order parameter), each dynamic pedestrian acts like a microscopic magnet that induces local polarization in the crowd, as demonstrated in section \ref{subsection5.1}. This phenomenon is characterized by an increase in observed dynamism and a concomitant reduction in symmetry, resulting in pronounced fluctuations in both velocity and density within localized regions. From the perspective of network dynamics, dynamic pedestrians can be analogized to active nodes. In a highly responsive structured network, such active perturbation generators will inevitably lead to a degradation of robustness.

\section{Advantages and limitations} \label{section6}

We have conducted extensive empirical validation and testing of the model. In this part, The most fundamental properties of the model will be discussed: how efficient the model is and under what circumstances the model may become ineffective.

\subsection{Property of linear time complexity} \label{subsection6.1}

\begin{figure}[ht!]
\centering
\includegraphics[scale=0.32]{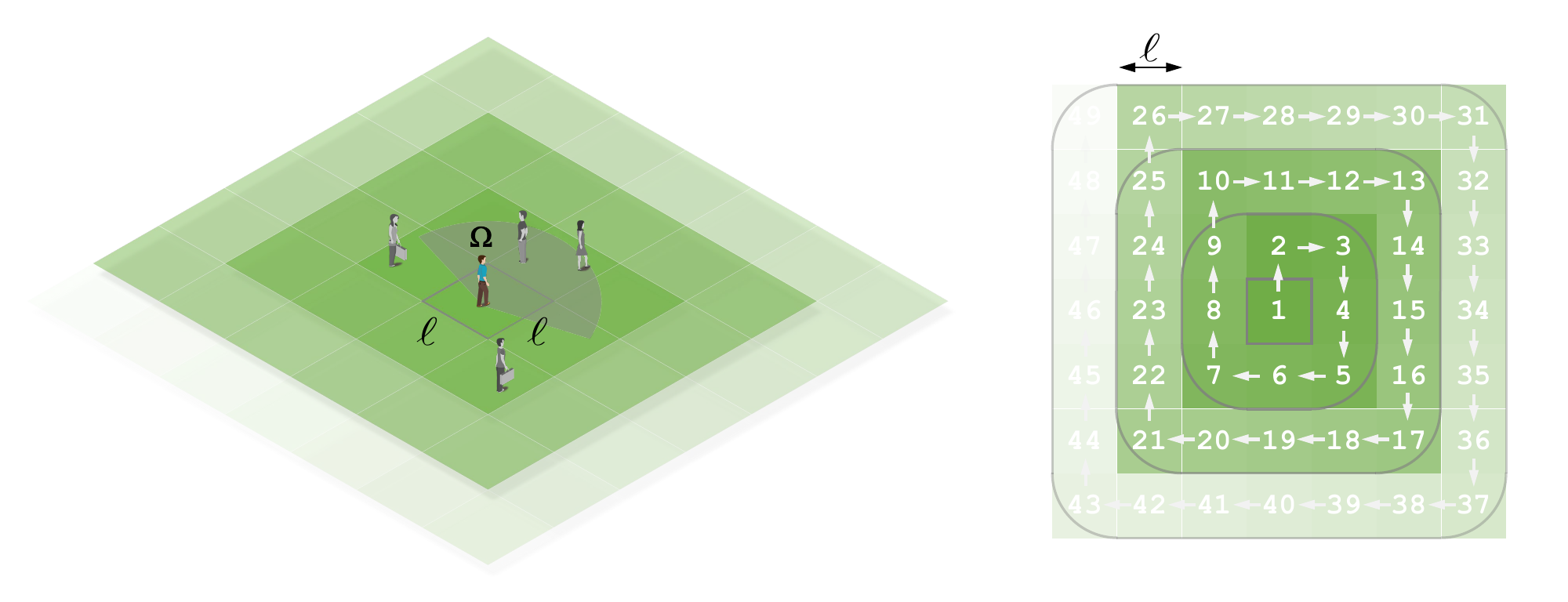}
\caption{Illustration of the memoryless clockwise depth-first search strategy for nearest neighbor.}
\label{fig20}
\end{figure}

Based on the method presented in section \ref{section3} (see Eq.\ref{10}), it is evident that the theoretical time complexity of this model ranges from $\mathcal{O}(n)$ (interactions without contact) to $\mathcal{O}(6n)$ (extreme case of a hexagonal-close-packed). The most challenging issue in modeling lies in the search of nearest-neighbor. While the wavefront method is inefficient in tackling this problem, we employ a memoryless-clockwise depth first search (DFS) strategy, which provides a fully regularized search process with higher efficiency and zero memory overhead. A schematic diagram of the search strategy is illustrated in Fig.\ref{fig20}, and the pseudocode representation is provided in Algo.\ref{MC-DFS}. On another note, the prerequisite for structured spatial search is determining the size of the structured unit. In our tests, we considered units of varying lengths. Given that nearest-neighbor distances vary with global density,  the following formula for calculating the adaptive unit length were also proposed:

\begin{equation}\label{20}
\ell = \left\lceil \frac{2}{{\sqrt {\rho \pi } }} \right\rceil 
\end{equation}

\begin{algorithm}
  \caption{Memoryless\_Clockwise\_DFS for Nearest Neighbor}\label{MC-DFS}
  \SetAlgoLined
  \KwData{initialization: \\
          \textit{direction\_search} = $(0, 1)$ \\
          \textit{agent\_i} $\in$ \textit{location\_grid} = $(x_0, y_0)$ \\
          $D_{ij} = \infty$ \\
          \textit{found} = \textit{false}}
  \KwResult{nearest neighbor \textit{j} and nearest neighbor relative distance $D_{ij}$}

  \textbf{Step 1: Outer Loop for Step Length}
  
  \For{$\textit{step\_length} = 1$ to $\infty$}{
      \If{\textit{found} == \textit{true}}{
          \textbf{return} {$j$ and $D_{ij}$}
      }

      \textbf{Step 2: Inner Loop for Rounds}
      
      \For{$\textit{round} = 1$ to $2$}{
          \If{\textit{round} == 2 \textbf{and} $\textit{step\_length} \bmod 2 == 1$ \textbf{and} $D_{ij} < (\textit{step\_length} - 1) \cdot \ell / 2$}{
              \textit{found} = \textit{true}
              \textbf{break}
          }

          \textbf{Step 3: Search within Step Length}
          
          \For{$\textit{step} = 1$ to $\textit{step\_length}$}{
              Search each agent in \textit{location\_grid}
              
              \If{$d_{ij} < D_{ij}$ \textbf{and} \textit{agent\_j} $\in \mathbf{\Omega}$}{
                  $D_{ij} = d_{ij}$} 
              Update \textit{location\_grid} $\mathrel{+}= \textit{direction\_search}$
          }

            \textbf{Step 4: Update Search Direction}
            
            \Switch{\textit{direction\_search}}{
                \Case{$(0, 1)$}{
                    \textit{direction\_search} = $(1, 0)$
                }
                \Case{$(1, 0)$}{
                    \textit{direction\_search} = $(0, -1)$
                }
                \Case{$(0, -1)$}{
                    \textit{direction\_search} = $(-1, 0)$
                }
                \Case{$(-1, 0)$}{
                    \textit{direction\_search} = $(0, 1)$
                }
            }
      }
  }
\end{algorithm}

\begin{figure}[ht!]
\centering
\includegraphics[scale=0.6]{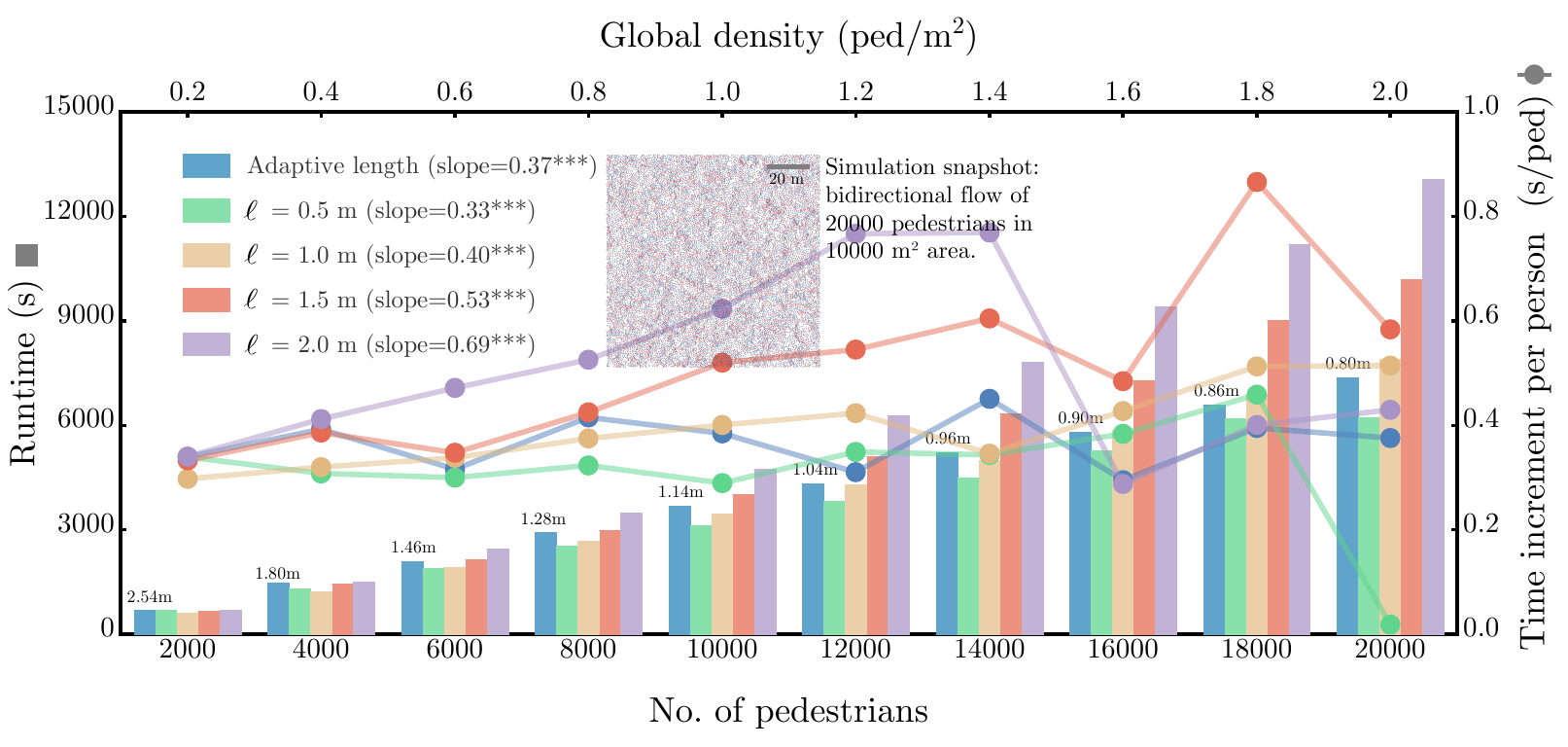}
\caption{Runtime for 3000 iterations under different basic unit lengths, showing its growth trend with the increase in the number of pedestrians. Clear linear relationships was observed, and *** showed significance at $p$ < 0.001 in ANOVA.}
\label{fig21}
\end{figure}

Benchmark tests were conducted on a laptop (Intel Core i7-11800H CPU, 16 GB memory) to evaluate simulation performance. The simulation tested five configurations for structured basic units: an adaptive length and four fixed lengths (0.5 m, 1 m, 1.5 m, and 2 m). In a 100\,m $\times$ 100\,m (10,000\,m\textsuperscript{2}) space, we randomly generated agents with equal increments (2,000 pedestrians per run) over 3,000 iterations. Fig.\ref{fig21} shows the results. The model clearly exhibited linear time complexity, with a statistically significant linear relationship between the basic unit length and the runtime. Comparisons across configurations revealed that larger unit sizes degraded computational efficiency, consistent with the linear fitting results. For adaptive unit length and fixed lengths of 0.5 m, 1 m, 1.5 m, and 2 m, the fitted slopes were 0.37, 0.33, 0.40, 0.53, and 0.69, respectively.

\subsection{Limitations} \label{subsection6.2}

\begin{figure}[ht!]
\centering
\includegraphics[scale=0.7]{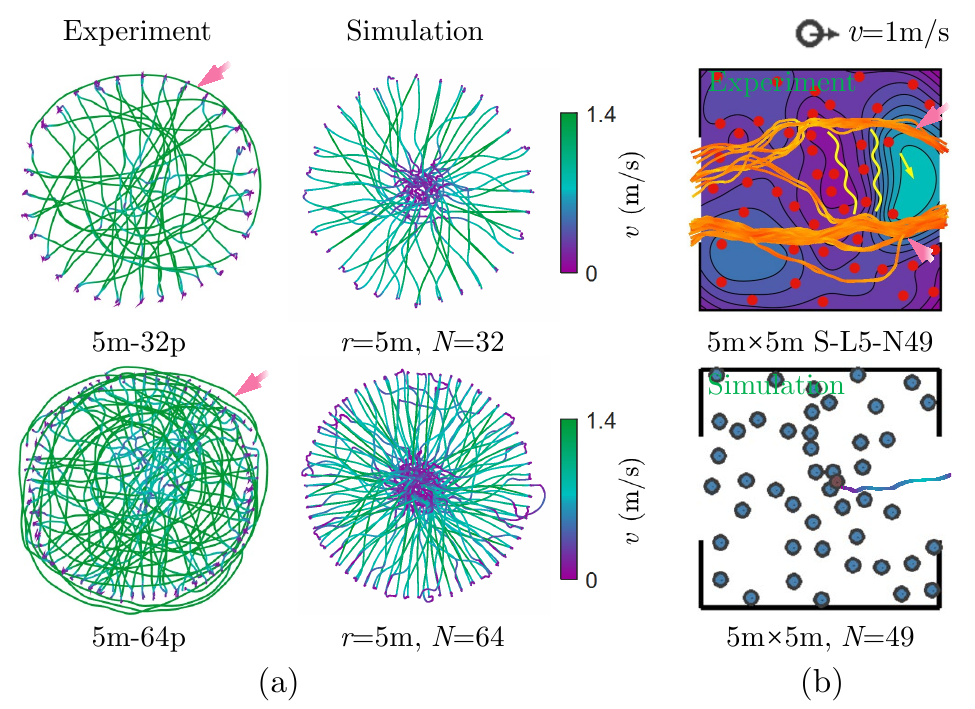}
\caption{Limitations of the model: comparison between experimental and simulation trajectories, the variable parameters set as \(\phi=\pi/2\) and \(\alpha=0.5\). (a) Antipode experiment \citep{xiao2019investigation}. (b) Crowd-cross experiment (static) \citep{wang2023exploring}.}
\label{fig22}
\end{figure}

We also explored the limitations of the model, which mainly include two aspects: \textbf{Intractability in task related with decision process.} As shown in Fig.\ref{fig22} (a) and (b), even simulating some local tactical-level behaviors, such as short-term route choice, is beyond the capability of the model, which is also a common limitation of most operational-level models. \textbf{Distortion in response to static environments.} In certain static environments, pedestrians may become trapped even when space is available. Within the framework of force-based models, there exist equilibrium positions in static environments that result in a balance of forces. These positions function analogously to black holes, attracting nearby pedestrians, as illustrated in Fig. \ref{fig22}(b).

\section{Conclusions} \label{section7}

In this paper, we propose a force-based general pedestrian model named CosForce, which consists of eight parameters and four equations. To simulate the anticipation and reaction behaviors of pedestrians, as well as the associated collision avoidance mechanisms, symmetric and asymmetric forces constrained by cosine functions are employed. Constructed under the principle of minimalism, the model can effectively captures general pedestrian motion patterns while offering the advantage of linear time complexity, making large-scale crowd simulations more tractable.

Empirical validation of the model was conducted for scenarios involving single-file motion and unidirectional flow. In simulations, we examined the phase separation process in crowds, including lane formation, stripe formation, and cross-channel formation, typically regarded as the phenomena of pedestrian self-organization. By analyzing entropy decrease and steady-state characteristics, quantitative observations clearly reveal the self-organization process. Furthermore, through mass gathering simulations, we quantitatively demonstrated the "catfish effect", a theoretical concept that currently lacks empirical validation. In this context, we analyzed polarization and the process of symmetry variation in crowds.

Finally, focusing on crowd collapse, we explored potential mechanisms through time series and flow-field analysis. Applying critical transition theory, we characterized crowd collapse as a transition process quantified by the order parameter. As crowd density increases, the first transition (second-order) occurs when the crowd becomes tightly compressed. At this stage, pedestrian interactions shift from self-propelled constraints of fundamental diagrams to granular like dynamics. The second transition (first-order) occurs during crowd collapse. At this stage, further compression approaches the limit, elevating the order parameter. When the crowd reaches a critical state (characterized by high density and a persistently rising order parameter), collapse initiates.

Given the simple rules and evaluation based on fundamental parameters (e.g., average normalized speed, order parameter), assessments of crowd dynamics can be intuitively derived, thus providing valuable insights and inspiration. However, since many findings are derived from  hindsight of simulations, their inherent limitations should not be overlooked. Furthermore, this model builds a bridge between traffic and pedestrian modeling on the basis of force-based methods.

\centerline{}
\section*{Data Availability}
Code and data can be found at: \url{https://drive.google.com/drive/folders/1gdkYmaUhcPgWp4i0Cffh2bph1fwvdQNp} (Google Drive) and a video summary is available at: \url{https://www.bilibili.com/video/BV17B1oYVEQm/}.

\centerline{}
\section*{Acknowledgments}
This work was supported by the National Natural Science Foundation of China (Grant No. 52072286, 71871189, 51604204), and the Fundamental Research Funds for the Central Universities (Grant No. 2022IVA108).

\appendix

\section{Experiments} \label{Appendix A}

The empirical validation in section \ref{section3} was conducted using data from experiment of unidirectional flow. The datasets are available at: \url{https://ped.fz-juelich.de/da/doku.php?id=start#data_section} (Pedestrian Dynamics Data Archive). Detailed experimental configurations and procedural specifics can be found in the works of \citet{cao2017fundamental}, as referenced. The specific details of the unidirectional pedestrian flow experiment are presented in Tab.\ref{table2}. Similar to the simulation analysis, data from frames 501-600 (corresponding to 4 s) were collected at a sampling frequency of 25 Hz for the purpose of empirical validation.

\begin{table}[htbp]
\centering
\caption{Runs with walking direction, widths, and number of candidates.}
\label{table2}
\begin{tabular}{lcccc}
\toprule
\textbf{Run name} & \textbf{Direction} & \textbf{Width b1 (m)} & \textbf{Width b2 (m)} & \textbf{Number of candidates} \\
\midrule
Uni\_corr\_500\_01 & Right to left & 1 & 5 & 148 \\
Uni\_corr\_500\_02 & Right to left & 2 & 5 & 760 \\
Uni\_corr\_500\_03 & Left to right & 5 & 3 & 916 \\
Uni\_corr\_500\_04 & Right to left & 4 & 5 & 909 \\
Uni\_corr\_500\_05 & Left to right & 5 & 5 & 905 \\
Uni\_corr\_500\_06 & Right to left & 4 & 5 & 913 \\
Uni\_corr\_500\_07 & Left to right & 5 & 3 & 914 \\
Uni\_corr\_500\_08 & Right to left & 2 & 5 & 477 \\
Uni\_corr\_500\_09 & Left to right & 5 & 1 & 310 \\
\bottomrule
\end{tabular}
\end{table}

\bibliographystyle{aasjournal}

\end{document}